\documentclass[format=acmsmall, review=false, screen=true]{acmart}

\usepackage{multirow}
\usepackage[normalem]{ulem}
\useunder{\uline}{\ul}{}

\usepackage{subcaption}
\usepackage{graphicx}
\usepackage{framed}
\usepackage{listings}   
\usepackage{xcolor}     
\usepackage{CJKutf8}
\usepackage{ragged2e}
\usepackage{balance}
\usepackage{color}

\acmJournal{TOSEM}
\usepackage{algorithmic}
\usepackage{textcomp}
\usepackage{booktabs}
\usepackage{bm}
\usepackage{float}
\usepackage{enumerate}
\usepackage{tcolorbox}
\usepackage{colortbl}
\usepackage{amsmath,amssymb,amsfonts,bm}
\usepackage{url}
\usepackage{rotating} 
\usepackage[table]{xcolor}
\usepackage{pgfmath}
\usepackage{colortbl}

\newcommand{\grayA}[1]{%
  \pgfmathsetmacro{\minval}{0.5}%
  \pgfmathsetmacro{\maxval}{2}%
  \pgfmathparse{max(min((#1 - \minval)/(\maxval - \minval),1),0)}%
  \pgfmathparse{1 - \pgfmathresult}%
  \xdef\grayval{\pgfmathresult}%
  \cellcolor[gray]{\grayval} #1%
}

\newtcolorbox{qoutebox}[3][]
{
  colframe=black!30!white,
  dcolback  = #2!10,
  #1,
}
\begin{document}

\title{Fine-Tuning Code Language Models to Detect Cross-Language Bugs}
\author{Zengyang Li}
\orcid{0000-0002-7258-993X}
\affiliation{%
  \institution{School of Computer Science, Central China Normal University}
  \city{Wuhan}
  \country{China}
}
\affiliation{%
  \institution{State Key Laboratory for Novel Software Technology, Nanjing University}
  \city{Nanjing}
  \country{China}
}
\email{zengyangli@ccnu.edu.cn}

\author{Yimeng Li}
\orcid{0009-0009-8509-4168}
\affiliation{%
  \institution{School of Computer Science, Central China Normal University}
  \city{Wuhan}
  \country{China}
}
\email{liyimengccnu@mails.ccnu.edu.cn}

\author{Binbin Huang}
\orcid{0009-0008-3763-5962}
\affiliation{%
  \institution{School of Computer Science, Central China Normal University}
  \city{Wuhan}
  \country{China}
}
\email{huangbinbin@mails.ccnu.edu.cn}

\author{Peng Liang}
\orcid{0000-0002-2056-5346}
\authornote{Corresponding author}
\affiliation{%
  \institution{School of Computer Science, Wuhan University}
  \city{Wuhan}
  \country{China}
}
\email{liangp@whu.edu.cn}

\author{Ran Mo}
\orcid{0000-0001-7556-153X}
\affiliation{%
  \institution{School of Computer Science, Central China Normal University}
  \city{Wuhan}
  \country{China}
}
\email{moran@ccnu.edu.cn}

\author{Hui Liu}
\orcid{0000-0001-7545-7986}
\affiliation{%
  \institution{School of Artificial Intelligence and Automation, Huazhong University of Science and Technology}
  \city{Wuhan}
  \country{China}
}
\email{hliu@hust.edu.cn}

\author{Yutao Ma}
\orcid{0000-0003-4239-2009}
\affiliation{%
  \institution{School of Computer Science, Central China Normal University}
  \city{Wuhan}
  \country{China}
}
\email{ytma@ccnu.edu.cn}


\renewcommand{\shortauthors}{Li et al.}

\begin{abstract}
With the rapid development of software engineering technologies, multilingual programming, i.e., developing a program using multiple programming languages (PLs) collaboratively, has become increasingly popular. This practice is often adopted to reuse existing code, to leverage advantages of specific PLs, to satisfy various software quality needs, and to enhance software development efficiency. However, it also introduces cross-language bugs (CLBs), bugs caused by interactions between multiple PLs. Most existing bug detection tools and techniques are primarily designed for a single PL and are less effective at detecting CLBs. Given the significant advancements of Pre-trained Language Models (PLMs) in the code domain, this work investigates the performance of pre-trained Code Language Models (CodeLMs) in CLB prediction tasks. To this end, we designed and developed a cross-language code identification tool (CLCFinder), and used it to collect and construct a CLB dataset, which involves combinations of three PLs (i.e., Python-C/C++, Java-C/C++, and Python-Java) with nine types of interaction mechanisms. We fine-tuned 13 popular CodeLMs in different sizes on this dataset, in which 80\% of the data was used for fine-tuning, 10\% for validation, and 10\% for testing. We then evaluated their performance in CLB detection. Additionally, we compared the fine-tuning performance of five top-performing models on current state-of-the-art (SOTA) single-language bug datasets to assess their application in CLB detection. Furthermore, we evaluated factors affecting the performance of fine-tuned CodeLMs in CLB detection, including the size of the fine-tuning dataset and the length of token sequence. Finally, we explored the impact of code comments on the performance of fine-tuned CodeLMs. The experimental results are that: First, all 13 CodeLMs exhibited varying degrees of performance improvement after fine-tuning. In particular, after fine-tuning, the UniXcoder-base model achieved the best performance, with an F1 score of 0.7407. Notably, within the scope of our experimental setup, small CodeLMs tended to achieve better performance than large CodeLMs, while large CodeLMs showed limited improvement. Second, the fine-tuned CodeLMs on single-language bug datasets performed poorly in CLB detection, further demonstrating the significant feature differences between CLBs and single-language bugs. Third, enlarging the fine-tuning dataset size significantly enhanced the model performance in CLB detection, but increasing token sequence length did not necessarily improve the model performance. Finally, regarding the impact of code comments, the performance varied significantly across different CodeLMs. Specifically, some CodeLMs' performance was improved when fine-tuned with code comments, while others showed degraded performance.
\end{abstract}




\maketitle

\section{Introduction}\label{chap:intro}

As modern software systems continue to grow in scale and complexity, developing software using multiple interacting programming languages (PLs) has become increasingly common in both industry and the open source community~\cite{delorey2007programming,capers2010software,li2021understanding,mayer2015empirical,tomassetti2014empirical}. By combining the strengths of different PLs, developers can build highly scalable and high performance software systems more efficiently. Since each PL has its own strengths and limitations, combining multiple PLs is often a necessary choice to achieve functionality or performance requirements that cannot be satisfied by a single-language alone~\cite{abidi2019behind,bissyande2013popularity,meyerovich2013empirical,vasilescu2013babel,yang2023demystifying}. As a result, multilingual software systems have gradually become the dominant form of modern software development compared to single-language systems~\cite{li2021understanding, mayer2015empirical, valverde2015punctuated}.

However, compared to single-language systems, multilingual software introduces significantly higher complexity. In multilingual software systems, interactions between different PLs are typically implemented through interfaces, libraries, or middleware~\cite{li2023understanding1}. For example, Python can interact with C/C++ through Cython or ctypes, while Java typically interacts with C/C++ through the Java Native Interface (JNI). Although these cross-language interaction mechanisms provide powerful functionality, they also introduce substantial system complexity~\cite{abidi2021multi}. Compared to single-language systems, the complexity of multilingual software does not only come from the simple aggregation of the complexity of individual languages. More importantly, it arises from the interactions between heterogeneous language components~\cite{abidi2021multi, bae2019towards, grichi2020impact, mayer2017multi, yang2024learning}. These cross-language interactions lead to a distinct class of bugs that caused originate from language interfaces and cross-language information flows, such as control flow and data flow. These bugs are known as cross-language bugs (CLBs)~\cite{lee2020broadening, li2022polycruise, li2022vulnerability}. Unlike bugs confined within a single-language, CLBs occur at language boundaries. Fundamental differences in execution models, memory management strategies, and type systems across languages make these bugs more hidden and harder to detect~\cite{mccormack2024study, li2023understanding2}. They also tend to lead to more severe consequences, including memory corruption, security vulnerabilities, and system crashes~\cite{yang2025dissecting}.

In this work, we distinguish between cross-language bugs and multilingual bugs, both existing in multilingual software projects.
A \textbf{cross-language bug} (CLB) refers to a bug that arise from direct interactions between different PLs (e.g., data flow and control flow interactions), typically facilitated through specific cross-language interaction mechanisms. JNI for Java and C/C++ as well as ctypes for Python and C/C++ are typical examples of cross-language interaction mechanisms. CLB examples include bugs occurring within Java’s JNI for Java and C/C++ cross-language interactions, or ctypes and Pybind11 for Python and C/C++ cross-language interactions. 
A \textbf{multilingual bug} refers to a bug whose fix involves the code change of multiple PLs, though the code of these PLs may not necessarily interact or interconnect. A multilingual bug example is a mismatch between frontend input validation written in JavaScript and backend processing in Python. Fixing the bug requires changes in both PLs, even though they do not directly interact.
It is important to note that CLBs are a specific subset of multilingual bugs. While both involve multiple PLs, CLBs are characterized by the presence of direct interactions between the involved languages through specific interaction mechanisms (such as JNI or ctypes). In contrast, multilingual bugs more broadly refer to issues requiring multi-language code changes, even if those languages do not directly interact.

Prior empirical studies on real-world CLBs show that these bugs are not random but arise from recurring root causes at language boundaries~\cite{yang2025dissecting}. Common causes include misuse of cross-language interfaces and violations of implicit API assumptions, mismatches in memory ownership and lifetime management between managed and unmanaged languages, and semantic inconsistencies in data representation and type conversion. In addition, differences in control flow and error-handling mechanisms across languages can lead to unexpected execution behaviors. CLBs are primarily interface- and interaction-centric, indicating that effective CLB analysis should focus on detecting anomalous cross-language interaction patterns rather than defects confined within a single-language.

Multilingual software may also contain bugs that exist only within a single-language component. Such bugs can usually be effectively handled using mature language specific testing and debugging tools~\cite{li2022vulnerability, li2023polyfuzz}. In contrast, CLBs represent a significant gap in modern software testing and bug diagnosis. And existing bug detection research mainly focuses on single-language scenarios. 
Current static analysis techniques usually rely on language specific intermediate representations and semantic rules, which makes it difficult for them to analyze  about cross-language data flow, control flow, and resource management in a unified way and often leads to high false positive rates~\cite{bae2019towards, lee2020broadening, park2023static, youn2023declarative, wei2018jn}.
Dynamic analysis techniques can achieve higher precision, but their effectiveness strongly depends on the availability and coverage of test cases. Real world multilingual systems often lack sufficient testing support~\cite{bai2018bridgetaint, li2022polycruise, li2023polyfuzz, xue2018ndroid}. In addition, generating high quality test cases typically requires substantial manual effort~\cite{hwang2021justgen}. Consequently, many CLBs cannot be effectively detected by existing static or dynamic techniques and may only surface during system execution. For example, an implicit data type mismatch between Python and a C extension might not trigger any compile-time warnings in either language but could lead to silent data corruption that is only observable under specific runtime conditions.

In recent years, bug detection approaches based on deep learning (DL) have attracted increasing attention~\cite{li2019deepfl, yang2022language, yang2024learning}. Unlike traditional static or dynamic analysis techniques, DL based methods are data driven and do not rely on explicitly encoded language rules or a unified intermediate representation. Since they avoid heavy language specific analysis, they are often easier to extend to multiple language combinations. In particular, rapidly advancing pre-trained code language models (CodeLMs), such as CodeBERT~\cite{feng2020codebert} and CodeLlama~\cite{roziere2023codellama}, have demonstrated strong capabilities in code understanding, code generation, and code completion tasks. These models can learn rich contextual and semantic representations from large scale code corpora.
As discussed above, CLBs are often caused by implicit and undocumented assumptions at language boundaries, such as mismatched memory ownership, inconsistent data representations, or incompatible lifecycle management across languages, which makes these recurring interaction patterns difficult to capture using rule-based analyses.
This allows them to implicitly capture recurring patterns in cross-language interactions, making them a promising technical foundation for CLB detection. However, how to effectively leverage CodeLMs for CLB detection remains an open research problem.

Although prior studies have begun to explore multilingual bug detection, existing work mainly focuses on bugs that require changes across multiple languages and does not sufficiently address bugs caused by direct cross-language interactions. For example, just-in-time multilingual bug prediction techniques~\cite{li2024exploratory} primarily target multilingual bugs rather than CLBs. In addition, in the CLB dataset constructed by Yang et al.~\cite{yang2024learning}, the Python code samples are bug free. As a result, models trained on this dataset cannot detect bugs in Python code, which limits their effectiveness in realistic cross-language scenarios. Based on these observations, it is necessary to conduct a systematic study of CLB detection and to comprehensively evaluate the effectiveness of CodeLM based approaches for this task.

This paper comprehensively investigates the problem of CLB detection and proposes a solution based on CodeLMs. Specifically, the \textbf{main contributions} of this work are summarized as follows:

\begin{itemize}
    \item[$\bullet$] We constructed a CLB dataset, which includes bugs from typical interaction patterns between mainstream PLs (Python, C/C++, and Java), providing a high-quality dataset for future research in CLB detection.
    \item[$\bullet$] We explored the performance of 13 popular CodeLMs in the task of CLB detection, evaluating their effectiveness for this task and revealing both the potential and limitations of these models in detecting CLBs.
    \item[$\bullet$] We investigated the factors that influence the performance of CodeLMs in the task of CLB detection, including training data volume and token sequence length. The findings provide empirical evidence for future research to optimize model design and improve CLB detection.
    \item[$\bullet$] We also explored the impact of code comments on the model's performance in detecting CLBs, evaluating the role of code comments in both the model's training and inference processes. Finally, we provided practical recommendations to developers based on our findings.
\end{itemize}

The remainder of this paper is organized as follows. Section \ref{chap:relat} discusses the related work; Section \ref{chap:case} describes the study design; Section \ref{chap:results} presents the study results; Section \ref{chap:discussion} interprets the study results with their implications; Section \ref{chap:threats} highlights the threats to the validity of the results; and Section \ref{chap:conclusions} concludes this work with future research directions.

\section{Related Work}\label{chap:relat}

\subsection{Software Bug Detection}\label{RelatedWork_A}

Software bug detection is a key task for ensuring software quality and has evolved from traditional static analysis and dynamic analysis to the current approaches leveraging machine learning and  CodeLMs~\cite{wang2024machine,li2024bug,li2024exploratory,mo2022exploratory}.

Common static analysis tools include Error Prone from Google~\cite{Google_ErrorProne}, Infer from Facebook~\cite{Facebook_Infer}, and SpotBugs~\cite{SpotBugs}, the successor to the widely used FindBugs tool~\cite{ayewah2008using}, are examples of popular static bug detection tools. Recent years, researchers have sought to enhance the comprehensiveness and accuracy of bug detection by evaluating and integrating different types of bug detection techniques, such as flow pattern matching~\cite{hovemeyer2004finding}, abstract interpretation~\cite{cousot1977abstract}, symbolic execution~\cite{king1976symbolic}, and program model checking~\cite{clarke2009model}.

Dynamic analysis detects bugs by monitoring and analyzing a program during its execution. Typical dynamic analysis techniques include dynamic symbolic execution~\cite{csallner2008dysy} and instrumentation~\cite{gosain2015survey}. The former combines symbolic execution with actual test paths to cover more program execution paths, while the latter uses an instrumentation mechanism to monitor the program's execution behavior.

With the rapid advancement of machine learning and deep learning techniques and their demonstrated effectiveness in various domains, researchers have increasingly explored their application in software bug detection, capitalizing on the growing volume of fault data generated during the software development process~\cite{liu2023semantic,deng2020software,li2017software,giray2023use}. Li et al. proposed a software bug prediction tool based on the RF algorithm, aimed at predicting whether code changes introduce bugs involving multiple PLs in a timely manner~\cite{li2024exploratory}. Wongpheng et al. utilized a multi-layer CNN to extract features from code snippets, achieving higher detection accuracy compared to traditional machine learning methods~\cite{wongpheng2020software}. Similarly, Khleel et al. employed a variant of RNN—Gated Recurrent Unit (GRU)—to capture temporal dependencies in code, combining it with CNNs for feature extraction, thereby improving detection performance~\cite{khleel2023novel}.

\subsection{Multilingual Bug and CLB Detection}\label{RelatedWork_B}

With the widespread use of multilingual software systems, an increasing number of studies have focused on bug detection and security analysis in multilingual software systems. Research indicates that the interaction mechanisms between PLs in multilingual software systems significantly impact software development, with the bug rate in cross-language dependencies being three times higher and the vulnerability rate being twice that of single-language dependencies~\cite{li2024multilingual,li2022vulnerability}. Similarly, in their study of cross-language dependencies, Grichi et al. found that cross-language dependencies are more likely to introduce bug and security vulnerabilities than single-language dependencies~\cite{grichi2020impact}. This highlights the importance of cross-language software analysis for software security, which has been actively studied. Sungjae Hwang et al. developed JUSTGen, a semi-automated tool, to study the behavior of JVM when handling JNI errors~\cite{hwang2024empirical,hwang2021justgen}. In addition, a decompilation method has been proposed for statically analyzing JNI programs, which effectively analyzes compiled JNI programs, overcoming the limitations of traditional static analysis tools~\cite{park2023static}. The PolyCrusie tool combines symbolic dependency analysis and data flow analysis to effectively identify cross-language security vulnerabilities, achieving significant results in Python-C programs~\cite{li2022polycruise}. xLoc uses a Transformer-based deep learning model to effectively detect and locate CLBs~\cite{yang2024learning}. Similarly, the MVD framework proposes an innovative approach by covering six major PLs and learning from a dataset containing over 11,000 real-world vulnerabilities, significantly improving vulnerability detection performance~\cite{zhang2024mvd}.  

\subsection{Large Language Models for Bug Detection}\label{RelatedWork_C}
The success of LLMs in natural language processing has led to growing efforts to apply them in software bug detection, resulting in a number of effective and well-evaluated approaches. In static analysis and bug detection, Li et al. proposed the LLIft framework, which combines LLMs with traditional static analysis techniques to enhance the detection of UBI (Use Before Initialization) bugs in the Linux kernel~\cite{li2024enhancing}. Similarly, GPTScan integrates LLMs with static analysis for detecting vulnerabilities in smart contracts~\cite{sun2024gptscan}. Khare et al. conducted a comprehensive analysis of LLMs' vulnerability detection capabilities and found that in some cases, LLMs perform similarly to static analysis tools like CodeQL~\cite{khare2023understanding}. Shu et al. evaluated PLMs and LLMs on multi-language vulnerability detection using over 30,000 real patches, demonstrating LLMs’ superior cross-language semantic understanding and high-risk vulnerability identification~\cite{shu2025large}.


\subsection{Conclusive Summary}\label{RelatedWork_D}
Although researchers have begun exploring cross-language software bugs, this field remains relatively underdeveloped. Existing studies mainly focus on specific PL combinations, such as Java and C/C++ (via JNI)~\cite{hwang2024empirical,hwang2021justgen,park2023static}, and xLoc can detect and location CLBs between Python and C~\cite{yang2024learning}. However, there are certain limitations in current research and tools, particularly in terms of dataset completeness. For example, xLoc's dataset lacks code for CLBs in Python, meaning that the tool can only detect CLBs in C~\cite{yang2024learning}. PolyFax is a toolkit for characterizing multilingual software. It can identify cross-language mechanisms within software, but it is not capable of determining whether a specific piece of code is cross-language code~\cite{li2022polyfax}. To address this issue, we designed and developed a cross-language code identification tool - CLCFinder, using it to collect and construct a CLB dataset from GitHub repositories, which includes Java-C/C++, Python-C/C++, and Python-Java combinations. Our goal is to advance CLBs detection techniques by constructing a more comprehensive and diverse CLB dataset.

In recent years, CodeLMs have demonstrated outstanding performance in software engineering tasks, including code understanding, code generation, and single-language bug detection~\cite{feng2020codebert,guo2020graphcodebert,qwen,hui2024qwen2,guo2024deepseek}. Based on this trend, we attempted to apply CodeLMs to CLB detection tasks. We developed CLCFinder and built a CLB dataset based on it. Using this dataset, we fine-tuned 13 prevailing CodeLMs and conducted a set of experiments, with the detailed experimental design presented in the following section
(Section \ref{chap:case}). 

\section{Study Design}\label{chap:case}

\subsection{Objective and Research Questions}\label{DesignRQ}
The \textbf{goal} of this study is to investigate the performance of the prevailing CodeLMs in detecting CLBs after fine-tuning, as well as the factors that influence their performance. 
We first fine-tuned several popular CodeLMs on our CLB dataset and evaluated their effectiveness on the cross-language detection task to validate the applicability and superiority of the dataset for this purpose. 
We then evaluated the performance of CodeLMs fine-tuned on mainstream single-language bug datasets when applied to cross-language scenarios, to further highlight the distinctiveness and necessity of our CLB dataset compared to existing single-language bug datasets. 
We subsequently designed a series of experiments to systematically analyze the factors affecting the models’ detection performance, given the varying performance of different CodeLMs in CLB detection and the fact that our dataset exhibits imbalances in the code length distribution while being significantly smaller in size compared to the pre-training datasets of the CodeLMs. Specifically, we examined the impact of the fine-tuning dataset size and the token sequence length on the performance of the fine-tuned CodeLMs in CLB detection.
Finally, we further investigated the impact of code comments — representing natural language elements — during fine-tuning, given that CodeLMs are pre-trained on both code and natural language texts, aiming to evaluate the effect of code comments on CLB detection performance.

To achieve the aforementioned goal, we formulated four research questions (RQs), which are described with their rationale: 
 
\textbf{RQ1: How do the CodeLMs perform after fine-tuning on our CLB dataset?}
While CodeLMs have demonstrated strong capabilities in software engineering tasks such as bug detection and code summarization~\cite{10172803}, their performance in cross-language scenarios remains largely unexplored. With this RQ, we aim to evaluate the effectiveness of fine-tuning a set of prevailing CodeLMs of varying sizes on our newly constructed CLB dataset. With this RQ, we can assess the ability of CodeLMs to detect CLBs and to analyze performance variations among models with different scales and architectures.

\textbf{RQ2: Can CodeLMs fine-tuned on the current SOTA single-language bug datasets detect CLBs? What is their performance?}
Prior research on bug detection has primarily relied on single-language bug datasets~\cite{guo2024comprehensive,yang2024learning}. However, in multilingual codebases, the transferability of such models is uncertain. With this RQ, we investigate whether CodeLMs fine-tuned on widely adopted single-language bug datasets can generalize to the CLB detection task. By comparing their performance with those trained directly on our CLB dataset, we aim to highlight the limitations of single-language training in cross-language contexts.

\textbf{RQ3: What factors influence the performance of fine-tuned CodeLMs in CLB detection?}
While large PLMs have shown promising results in various coding tasks, their effectiveness can be sensitive to different factors during fine-tuning, especially in low-resource or imbalanced data settings~\cite{wang2022no}. In this RQ, we investigate how two key factors — the size of the fine-tuning dataset and the length of input token sequences - affect the performance of fine-tuned CodeLMs in detecting CLBs. First, we vary the amount of training data to simulate low-resource scenarios and evaluate how model performance changes with increasing data availability. Second, we analyze the effect of token sequence length constraints, which directly determine the model's capacity to capture complete code structures and contextual dependencies, on the bug detection capability. Understanding the impact of these factors can help guide future dataset construction and model design in CLB detection.

\textbf{RQ4: How do code comments affect the performance of fine-tuned CodeLMs in detecting CLBs?}
Many CodeLMs are pre-trained using corpora that include both code and natural language documentation (e.g., comments, docstrings), which enables them to capture semantic information beyond the syntax of source code~\cite{wang2021codet5,feng2020codebert,guo2022unixcoder,hui2024qwen2}. Code comments often provide valuable contextual information that can help clarify code intent, logic, or constraints, especially useful in understanding buggy behaviors. In this RQ, we examine whether incorporating code comments during fine-tuning enhances model performance in the CLB detection task. We compare model performance across two settings: one where code comments are preserved and the other where code comments are removed. This comparison enables us to assess whether code comments enhance the model’s ability to understand and detect CLBs by providing additional contextual information, or if code comments instead introduce irrelevant information that degrade detection performance in a cross-language context. The findings of this RQ could inform best practices for dataset preprocessing and preparation in developing cross-language CodeLMs.

\subsection{Cross-Language Code Identification Tool}\label{tool_1}

Currently, there have been no tools for identifying cross-language code. To fill this gap, we developed CLCFinder specifically designed for the interaction characteristics of Python-C/C++, Java-C/C++, and Python-Java language combinations. CLCFinder is included in the replication package of this work to support reproducibility. We used CLCFinder to collect cross-language code data. As shown in Table \ref{table:cross-language_mechanisms}, it can identify nine cross-language interaction mechanisms across Python, C/C++, and Java, including six between Python and C, one between Python and Java, and two between Java and C/C++.

The selection of these nine cross-language interaction mechanisms follows the principles of representativeness and practical feasibility. Python, C/C++, and Java have long been among the most widely used PLs according to GitHub statistics (GitHub Octoverse \cite{github_octoverse2025}) and language popularity rankings such as TIOBE~\cite{tiobe_index}, and they frequently coexist in real-world open-source projects. Motivated by this observation, we surveyed and analyzed the common pairwise cross-language interaction mechanisms among these three languages. For each interaction mechanism, we conducted an in-depth analysis of its characteristics and designed corresponding code analysis logic, which was then implemented in the CLCFinder tool. The final set of nine mechanisms was determined based on extensive investigation of documentation, open-source codebases, and real-world projects, revealing that these mechanisms are widely used in practice and exhibit relatively stable implementation patterns, making them amenable to automated identification. It is worth noting that some cross-language interaction mechanisms allow flexible or diverse implementations; therefore, CLCFinder may not capture all possible variants. However, our design prioritizes the correctness of identified results, ensuring that the collected samples indeed involve genuine cross-language code interactions.

\begin{table}[h]
\centering
\renewcommand\arraystretch{1.2}
\caption{Cross-language interaction mechanisms supported in CLCFinder}
\resizebox{0.7\textwidth}{!}{
\begin{tabular}{ll}
\hline
\textbf{Language Pair}                     & \textbf{Cross-Language Interaction Mechanisms} \\ \hline
Python-C/C++ & PythonC, ctypes, Boost.Python, Cffi, SWIG, Pybind11              \\ \hline
Java-C/C++  & JNI, JNA                   \\ \hline
Java-Python                   & Jython                \\ \hline
\end{tabular}
}
\label{table:cross-language_mechanisms}
\end{table}

To illustrate the working principle of CLCFinder, we use the JNI examples shown in Listings \ref{lst:jni1_1}, \ref{lst:jni1_2}, \ref{lst:jni2_1}, and \ref{lst:jni2_2}.
In the JNI project directory, there are three files: \textsc{NativeMethod.java}, \textsc{NativeCaller.java}, and \textsc{NativeMethod.c}.
As shown in Listings \ref{lst:jni1_1} and \ref{lst:jni1_2}, the \texttt{NativeMethod} class declares a native method (i.e., \texttt{sayHello()}) and loads the native code library to invoke C code. \textsc{NativeMethod.c} contains the C code being called (the same applies to C++ code). The \texttt{NativeCaller} class instantiates \texttt{NativeMethod} and calls method \texttt{sayHello()}, thereby triggering a call to the C library code.
We consider Line 3 in Listing \ref{lst:jni1_2} as Java code that calls a function written in C. If a fault or error occurs on this line, we classify the bug on this line as a CLB. Therefore, we categorize the code in Line 3 as a cross-language code line.

\begin{lstlisting}[language=Java, caption={NativeMethod.java file of the first JNI example},label={lst:jni1_1}]% 设置语言和标题

    public class NativeMethod {
        //Native method declaration
        public native void sayHello();
    
        // Load local C library
        static {
            System.loadLibrary("nativeMethod");
        }
    }
\end{lstlisting}

\begin{lstlisting}[language=Java, caption={NativeCaller.java of the first JNI example},label={lst:jni1_2}]% 设置语言和标题
    
    public class NativeCaller {
        public static void main(String[] args) {
            NativeMethod nativeMethod = new NativeMethod();
            nativeMethod.sayHello(); 
        }
    }
\end{lstlisting}

Another example is presented in Listings \ref{lst:jni2_1} and \ref{lst:jni2_2}: if the called C code has a return value, and we assign this return value to variable \texttt{var\_a} (Line 5 in Listing \ref{lst:jni2_2}), which is then referenced by variable \texttt{var\_b}, and \texttt{var\_c} subsequently depends on \texttt{var\_b} (i.e., the data flow from \texttt{var\_a} to \texttt{var\_b} and then to \texttt{var\_c}), a fault occurring during this data transfer process would also be categorized as a CLB. However, theoretically, data can be passed through an unlimited number of variables. When data pass through numerous variables, any bug occurring during transmission is not necessarily due to the cross-language code invocation. Thus, we do not classify it as a CLB, nor do we consider the corresponding code lines as cross-language code line. Therefore, we only examine cases where data are transferred a maximum of three times—starting from receiving data \texttt{var\_a} from a cross-language function call, passing it to \texttt{var\_b}, and then to \texttt{var\_c}. And \texttt{var\_d} in the code is not considered as cross-language code. The rationale behind this constraint is that, as the number of data transfers increases, the dependence of the data on the original cross-language interaction gradually weakens, and any resulting fault is more likely to stem from intra-language logic rather than the cross-language interface itself. In addition, limiting the number of transfers ensures coverage of typical cross-language data flow scenarios while reducing analysis complexity and minimizing false positives. We only categorize code lines involving \texttt{var\_a}, \texttt{var\_b}, and \texttt{var\_c} as cross-language code (\texttt{var\_a}, \texttt{var\_b}, and  \texttt{var\_c} can also be analogized as functions), and any faults in these lines are considered CLBs. Specifically, we categorize the function that contains the cross-language code line(s) as the cross-language function. 

\noindent
\begin{minipage}{\textwidth}
\begin{lstlisting}[language=Java, caption={NativeMethod.java of the second JNI example},label={lst:jni2_1}, aboveskip=10pt] % 设置语言和标题

    public class NativeMethod {
        //Native method declaration
        public native int getValue();
    
        // Load local C library
        static {
            System.loadLibrary("nativeMethod");
        }
    }

\end{lstlisting}
\end{minipage}

\noindent
\begin{minipage}{\textwidth}
\begin{lstlisting}[language=Java, caption={NativeCaller.java of the second JNI example},label={lst:jni2_2}] % 设置语言和标题

    public class NativeCaller {
        public static void main(String[] args) {
            NativeMethod nativeMethod = new NativeMethod();
            x = ...
            ...
            var_a = nativeMethod.getValue(); 
            ...
            var_b = x / var_a;
            ...
            var_c = var_c + var_b;
            ...
            var_d = func(var_c);
            ...
        }
    }
\end{lstlisting}
\end{minipage}

Similarly, we conducted an in-depth study of various cross-language interaction mechanisms to enable the recognition of cross-language code between the pairs of common PLs, specifically Python, C/C++, and Java. The detailed information is available in Table \ref{table:cross-language_mechanisms}. When we study these cross-language interaction mechanisms, we found that the interactions between Python or Java and C/C++, the C/C++ code is typically on the called side, often in the form of library files, such as \texttt{.so} files of Linux and \texttt{.dll} files of Windows rather than directly interacting with C/C++ code.  Therefore, to streamline tool implementation and improve data collection efficiency, we limit our detection to instances where Python and Java code invoke C/C++ libraries (i.e., only identifying cross-language code in cases where Python and Java interact with C/C++).

\subsection{Data Collection}\label{data_collection}

\begin{figure*}
    \centering
    \includegraphics[width=1\linewidth]{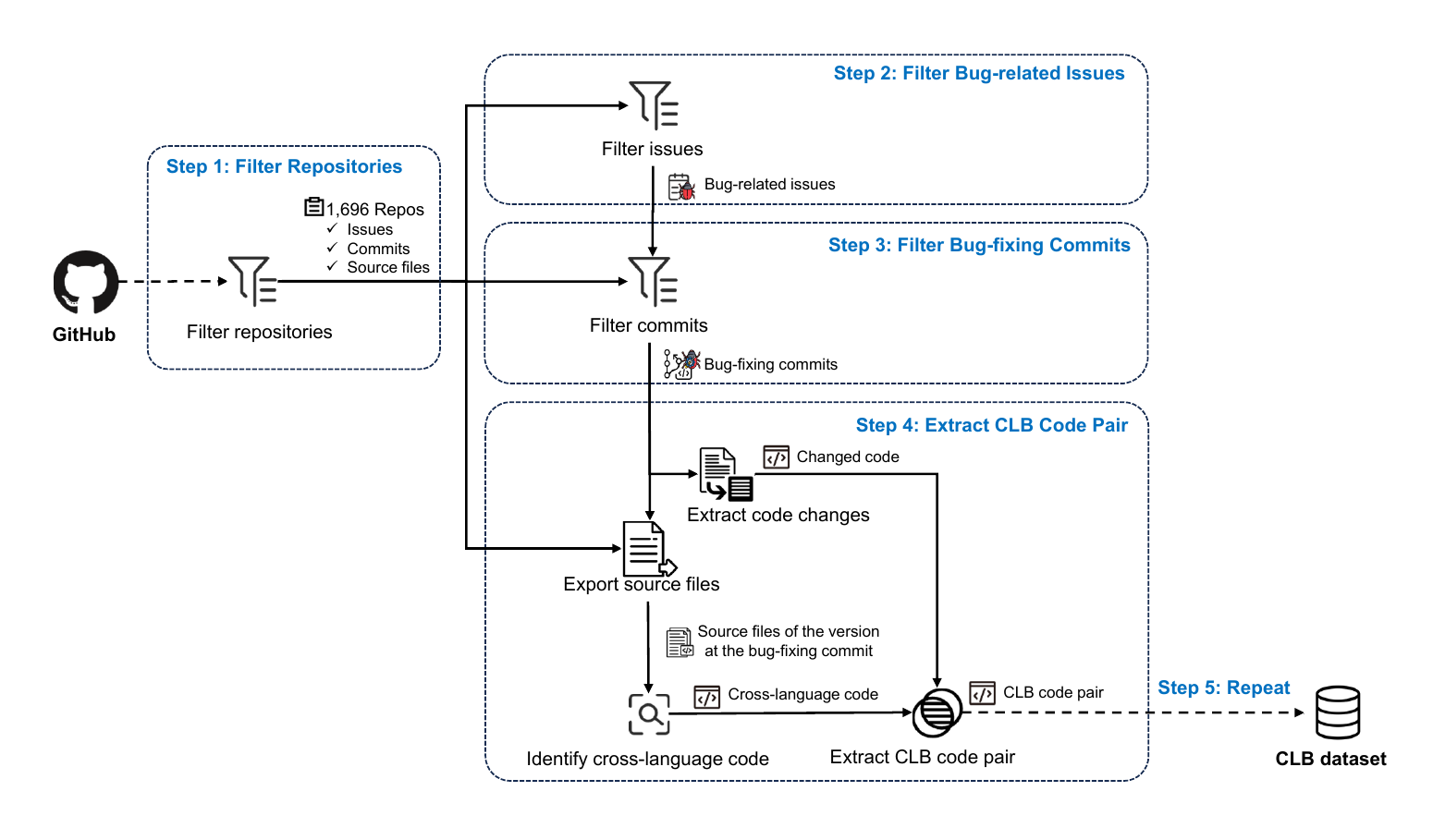}
    \caption{Data collection process}
    \label{fig:data_collect}
\end{figure*}

The data collection process is depicted in Figure~\ref{fig:data_collect}. To automate the data collection process, we developed a data collection tool that automatically retrieves relevant data from GitHub based on specified criteria. The detailed steps of our data collection process are described as follows:

\textbf{Step 1}: \textit{Filter Repositories}. We used the GitHub API to retrieve information on repositories that meet a set of specific predefined criteria. To balance sample size with project representativeness, we established a lower bound for the number of stars a repository must have, ultimately setting this threshold at 500 based on iterative experimentation. This approach ensured the inclusion of sufficiently popular and active projects, thereby enhancing the reliability and generalizability of our analysis. We focused on repositories involving cross-language collaboration, specifically those that include combinations of Java and C/C++, Python and C/C++, or Java and Python. To ensure that each language plays a substantive role in the project, we required that the proportion of code written in each PL exceeds 5\%. This constraint helps exclude PLs used only for peripheral tasks (e.g., testing or build scripts), thereby enhancing the focus on genuine cross-language development practices.
We saved the information for repositories meeting these criteria and manually reviewed each repository's description to filter out non-software repositories, such as those used for educational resources or personal static websites. Ultimately, we collected 1,696 repositories that were verified and used for subsequent analysis.

\textbf{Step 2}: \textit{Filter Bug-related Issues}. Given a repository from the collection of Step 1, we retrieved all issue information from that repository and filtered issues based on specific criteria. Specifically, we required the issue status to be closed, ensuring that the issue had been addressed or resolved. In addition, the issue's label or description (including the title and detail information) needed to indicate relevance to bugs. We used keyword matching in each issue's label or description, marking issues as bug-related if keywords associated with bugs were found. Through this process, we collected all bug-related issues for each repository.

\textbf{Step 3}: \textit{Filter Bug-fixing Commits}. We retrieved all commit information from the repository, including SHA, description and title of each commit. We then searched the commit titles for references to the issue IDs obtained in Step 2, using regular expressions to match the patterns. Developers typically include the issue ID in the commit title to indicate that the commit addresses the problem outlined in that issue. Since the issue is bug-related, we identified the corresponding commit as a bug-fixing commit. As a result, we collected all of the commits associated with bug fixes.

\textbf{Step 4}: \textit{Extract Bug Code}. We first cloned each repository to the local environment. For all bug-related commits identified in Step 3, we checked out the repository code to the specific commit version using SHA of the commit and exported the relevant source files. We then applied CLCFinder (see Section \ref{tool_1} for details) to identify cross-language code within the exported source files. With ctags~\cite{universal_ctags}, we extracted the functions containing cross-language code lines, referring to these functions as cross-language functions. Next, we extracted code changes from the bug-fixing commits. If a commit modified any cross-language function, we considered this commit to have fixed a CLB within that function, and thus labeled the function as a CLB-fixing function. We then retrieved the parent commit of the current commit and extracted the version of the function prior to modification (discarding newly added functions, as they lack a prior version), which represents the version containing the CLB. At this point, we had collected a pair consisting of the buggy cross-language function and its fixed version (i.e., clean function). And these functions construct a CLB code pair. We repeated this process for all commits in the repository, thereby collecting the complete set of CLB code pairs.

\textbf{Step 5}: \textit{Repeat}. We repeated Step 2, 3, and 4 for each repository identified in Step 1, resulting in the complete CLB dataset.

\subsection{Selection of CodeLMs}\label{CodeLMs}

Our research goal is to fine-tune CodeLMs on our CLB dataset to detect CLBs. The following are the criteria we employed to select the CodeLMs:
Firstly, since we need to fine-tune the CodeLMs, they must be open-source. Open-source CodeLMs ensure the transparency and reproducibility of the research, as well as provide more opportunities for customization and optimization. Secondly, the CodeLMs need to support binary classification. We treat the code bug prediction task as a binary classification problem, i.e., determining whether a code snippet contains a bug. Thirdly, large pre-trained models typically require significant computational resources (such as GPUs or TPUs). Therefore, the CodeLMs that we choose must balance good performance and reasonable resource consumption, not beyond our computational capacity. Lastly, the model should be one of the popular or latest CodeLMs, ensuring that the CodeLMs we use are competitive in code understanding and bug detection tasks.
Based on these criteria, we selected 13 CodeLMs for our experiments. Table \ref{table:details_CodeLMs} shows the detailed information about these models. These models have demonstrated outstanding performance in tasks such as code understanding, code generation, code search, and code completion~\cite{feng2020codebert,guo2020graphcodebert,qwen,hui2024qwen2,guo2024deepseek}.

\begin{table*}[ht]
\centering
\renewcommand\arraystretch{1.2}
\caption{Details of the CodeLMs studied in our work}
\resizebox{\textwidth}{!}{ 
\begin{tabular}{lllll}
\hline
\textbf{Model} & \textbf{Size} & \textbf{Context Length} & \textbf{Pre-training Dataset} & \textbf{Architecture} \\ \hline
CodeBERT-base~\cite{feng2020codebert} & 125M & 512 & CodeSearchNet~\cite{husain2019codesearchnet} & \multirow{3}{*}{Encoder-only} \\
UniXcoder-base~\cite{guo2022unixcoder} & 125M & 768 & C4~\cite{raffel2020exploring}, CodeSearchNet & \\
GraphCodeBERT-base~\cite{guo2020graphcodebert} & 125M & 512 & CodeSearchNet & \\
\cline{1-5}
NatGen~\cite{chakraborty2022natgen} & 220M & 512 & GitHub & \multirow{4}{*}{Encoder-decoder} \\
CodeT5-base~\cite{wang2021codet5} & 220M & 512 & CodeSearchNet, BigQuery~\cite{google_bigquery_github} & \\
CodeT5p-220M~\cite{wang2023codet5+} & 220M & 512 & codeparrot/github-code~\cite{codeparrot_github_code} & \\
CodeT5p-770M~\cite{wang2023codet5+} & 770M & 512 & codeparrot/github-code & \\
\cline{1-5}
Deepseek-Coder-1.3B-base~\cite{guo2024deepseek} & 1.3B & 16384 & $^*$ & \multirow{6}{*}{Decoder-only} \\
StarCoder2-3B~\cite{lozhkov2024starcoder} & 3B & 16384 & The Stack v2 & \\
Granite-3B-Code-base-2k~\cite{mishra2024granite} & 3B & 2048 & GitHub Code Clean~\cite{codeparrot_github_code_clean}, Starcoder data~\cite{lozhkov2024starcoder} & \\
Code\_Qwen1.5-7B~\cite{qwen} & 7B & 65536 & $^*$ & \\
Qwen2.5-Coder-7B~\cite{hui2024qwen2} & 7B & 32768 & $^*$ & \\
CodeLlama-7b-hf~\cite{roziere2023codellama} & 7B & 16384 & $^*$ & \\ \hline
\end{tabular}
}%

    \begin{minipage}{\linewidth}
        \vspace{0.2cm}
        \footnotesize{$^*$ There is no specific information about the pre-training dataset.}
    \end{minipage}
    
\label{table:details_CodeLMs}
\end{table*}

\subsection{Evaluation Metrics}\label{valuation_metrics}

In this study, we use a binary classification method to detect CLBs, determining whether a sample contains a bug (label = 1 or 0). We evaluate the performance of CodeLMs using several classification metrics, including accuracy (Equation \ref{eq:accuracy}), precision (Equation \ref{eq:precision}), recall (Equation \ref{eq:recall}), F1 score (Equation \ref{eq:f1}), and AUC (Area Under Curve, Equation \ref{eq:auc}). These metrics provide a comprehensive evaluation of the CodeLMs' strengths and weaknesses, ensuring that the evaluation results are both scientifically robust and practically valuable. 

\begin{equation}
    Accuracy=\frac{TP+TN}{TP+TN+FP+FN}\label{eq:accuracy}
\end{equation}

\begin{equation}
Precision=\frac{TP}{TP+FP}\label{eq:precision}
\end{equation}

\begin{equation}
Recall=\frac{TP}{TP+FN}\label{eq:recall}
\end{equation}

\begin{equation}
F1=2\times\frac{Precision\times Recall}{Precision+Recall}\label{eq:f1}
\end{equation}

\begin{equation}
AUC=\int_{0}^{1} TPR(FPR) \, d(FPR)\label{eq:auc},
\end{equation}

\noindent where \textbf{TP}, \textbf{TN}, \textbf{FP}, \textbf{FN} represent true positives, true negatives, false positives, and false negatives, respectively. The \textbf{TPR} (True Positive Rate), also known as Recall, is calculated as $\frac{TP}{TP+FN}$, while the \textbf{FPR} (False Positive Rate) is calculated as $\frac{FP}{FP+TN}$.

\subsection{Fine-tuning Environments and Methods for CodeLMs}\label{env}

\begin{table}[h]
\renewcommand\arraystretch{1.2}
\caption{Fine-tuning environments and methods for the CodeLMs}
\resizebox{\textwidth}{!}{
\begin{tabular}{lll}
\hline
\textbf{Environment} & \textbf{CodeLM} & \textbf{Fine-tuning Method} \\ \hline
\multirow{4}{*}{\begin{tabular}[c]{@{}l@{}}NVIDIA GeForce \\ RTX 4090 24G\end{tabular}} & CodeBERT-base, Graph CodeBERT-base, UniXcoder, & \multirow{2}{*}{Full fine-tuning} \\
 & CodeT5-base, CodeT5p-220M, CodeT5P-770M, NatGen &  \\ \cline{2-3} 
 & Granite-3B-Code-base-2k, StarCoder2-3B, CodeQWen1.5-7B, & \multirow{2}{*}{LoRA} \\
 & Qwen2.5\_Coder-7B, CodeLlama-7b-hf &  \\ \hline
NVIDIA L20 48G & Deepseek-Coder-1.3B-base & LoRA \\ \hline
\end{tabular}
}
\label{tab:env}
\end{table}

For different models, we adopted different fine-tuning environments. For small models, we performed full fine-tuning. However, due to our limited GPU resources, for large models with more parameters, we employed the LoRA (Low-Rank Adaptation) technique~\cite{hu2021lora}, which requires less computational resources and offers higher efficiency compared to full fine-tuning. The fine-tuning environment for each model is shown in Table \ref{tab:env}.

\subsection{Experimental Design for Each Research Question}\label{chap:expdesign}


Before formally introducing the experimental procedure, we first clarify that in all experimental settings of this study, the input context length of the CodeLMs is uniformly limited to no more than 512 tokens, except for the token sequence related experiments in RQ3, where different maximum token lengths are explicitly evaluated. The rationale for this setting can be summarized as follows.
(1) This maximum token length is mainly constrained by the hardware resources used in the experiments, especially the available GPU memory. 
For models based on the Transformer architecture, the memory and computation cost of the self attention mechanism increases approximately with the square of the input sequence length [\textcolor{violet}{73}]. 
(2) From the perspective of data distribution, this length limitation is reasonable. As shown in Figure \textcolor{violet}{4}, statistics based on the tokenizers of some CodeLMs indicate that most samples (approximately 75 percent) contain no more than 512 tokens after tokenization. Therefore, this limitation does not significantly affect the information completeness of most samples.
(3) This setting is also aligned with the architectural constraints of the evaluated models, as 6 out of the 13 CodeLMs used in this study support a maximum context length of 512 tokens (see Table \textcolor{violet}{2}).

In the implementation process, the original input is first tokenized. If the generated token sequence length exceeds the predefined upper limit of 512, only the first 512 tokens are kept and the remaining tokens are discarded. If the sequence length is less than or equal to 512, the entire sequence is preserved and fed into the model. This fixed length truncation strategy effectively controls the input size of the model and avoids GPU memory overflow caused by excessively long sequences.

\subsubsection{RQ1: Experimental Setup for Baseline and Fine-Tuning on CLB Dataset}
In the RQ1 experiment, we first randomly split the CLB dataset into training, validation, and test sets with a ratio of 8:1:1. We then loaded 13 pre-trained CodeLMs using the \texttt{AutoModelForSequenceClassification} interface and adapted all of them to a binary classification task so that the models can produce classification results. 

\begin{table}[t]
\centering
\caption{Fine-tuning Hyperparameter Settings for Different CodeLMs}
\label{tab:hyperparameters}

{
\begin{tabular}{lcccccc} 
\toprule
\textbf{Model} &
\textbf{Acc. Steps} &
\textbf{Batch} &
\textbf{Epochs} &
\textbf{LR} &
\textbf{WD} &
\textbf{Warmup}  \\
\midrule
CodeBERT-base        & 4  & 32 & 20 & $1\times10^{-4}$ & $1\times10^{-2}$ & 0.10 \\
GraphCodeBERT-base   & 4  & 32 & 20 & $1\times10^{-4}$ & $1\times10^{-2}$ & 0.10 \\
UniXcoder-base       & 4  & 32 & 20 & $1\times10^{-4}$ & $1\times10^{-2}$ & 0.10 \\
CodeT5-base          & 16 & 8  & 20 & $1\times10^{-4}$ & $1\times10^{-2}$ & 0.10 \\
CodeT5p-220M         & 16 & 8  & 20 & $1\times10^{-4}$ & $1\times10^{-2}$ & 0.10 \\
CodeT5p-770M         & 32 & 4  & 10 & $1\times10^{-4}$ & $1\times10^{-2}$ & 0.05 \\
NatGen               & 16 & 8  & 20 & $1\times10^{-4}$ & $1\times10^{-2}$ & 0.10 \\
DeepSeek-Coder-1.3B-base & 16 & 8  & 10 & $5\times10^{-5}$ & $1\times10^{-2}$ & 0.06 \\
Granite-3B-Code-base-2k & 16 & 4  & 15 & $3\times10^{-4}$ & $1\times10^{-2}$ & 0.06 \\
StarCoder2-3B        & 16 & 4  & 10 & $3\times10^{-4}$ & $1\times10^{-2}$ & 0.05 \\
CodeQwen1.5-7B       & 64 & 1  & 10 & $3\times10^{-4}$ & $1\times10^{-2}$ & 0.06 \\
Qwen2.5\_Coder-7B    & 64 & 1  & 10 & $3\times10^{-4}$ & $1\times10^{-2}$ & 0.06 \\
CodeLLama-7b-hf      & 4  & 24 & 10 & $1\times10^{-4}$ & $1\times10^{-2}$ & 0.06 \\
\bottomrule
\end{tabular}

\smallskip
\begin{minipage}{\linewidth}
\small
\textit{Note:} Acc. Steps = Gradient Accumulation Steps;  
Batch = Per-device Batch Size;
Epochs = Epochs Size;
LR = Learning Rate;  
WD = Weight Decay;
Warmup = Warmup Ratio.
\end{minipage}
}
\end{table}

In addition, we introduced a general LLM as an inference-based reference baseline for comparison. Considering the tradeoff between performance and inference cost, we selected GPT-4o-mini as the baseline model. This model is only used for inference on the test set and is not involved in any training or fine-tuning process. The prompt used for inference is shown in Table \ref{tab:prompt}.

Next, we fine-tuned the 13 CodeLMs on the CLB dataset. The training set is used for model training, and the validation set is used for model selection and hyperparameter tuning. After fine-tuning, we evaluated the fine-tuned models on the test set to assess their CLB detection performance. The hyperparameters used for fine-tuning are shown in Table \ref{tab:hyperparameters}. These hyperparameters are chosen after many experiments and adjustments to ensure stable and representative performance for all CodeLMs. All CodeLMs are trained using the AdamW~\cite{loshchilov2017decoupled} optimizer with a linear learning rate scheduler and a warmup strategy. The specific learning rate schedule and warmup settings vary across CodeLMs, and the details are reported in Table \ref{tab:hyperparameters}.
It is important to note that the hyperparameters used for each model are kept consistent across all experiments in this study. Specifically, for RQ1, RQ2, RQ3, and RQ4, the same model always adopts an identical set of hyperparameters. This design avoids potential interference caused by hyperparameter variations and ensures fair and comparable results across different experiments.

\begin{table}[htbp]
\caption{GPT-4o-mini Prompt Used for CLB Detection}
\label{tab:prompt}
\centering
{
\begin{tabular}{lp{10cm}}
\hline
\textbf{Role} & \textbf{Prompt} \\
\hline
System &
You are an expert software bug detection system specialized in cross-language bugs. \par

A cross-language bug refers to a defect that occurs at the interaction boundary between two different programming languages (e.g., JNI, ctypes, FFI, native extensions), caused by incorrect assumptions or mismatches across languages. These bugs typically involve, but are not limited to: \par
\begin{itemize}
  \item Incorrect data type mapping or size mismatch
  \item Memory management issues across language boundaries
  \item Improper pointer or reference handling
  \item Incorrect function signatures or calling conventions
  \item Resource lifecycle mismatches between languages
\end{itemize}

Given the following code snippet, determine whether it contains a cross-language bug. \par

Focus only on bugs that originate from or are triggered by cross-language interactions. Do NOT consider purely single-language bugs. \par

Respond with exactly one word:
\begin{itemize}
  \item YES if the code contains a cross-language bug
  \item NO if it does not
\end{itemize}

Do not provide explanations or any additional text.
\\
\hline
User &
Please analyze this function code for cross-language bugs: \par
\texttt{\{code\}}
\\
\hline
\end{tabular}
}
\end{table}

\subsubsection{RQ2: Experimental Setup for Transfer from Single-Language Bug Datasets}

To address RQ2, we fine-tuned five CodeLMs (i.e., CodeBERT-base, GraphCodeBERT-base, UniXcoder-base, CodeT5p-220M, NatGen) that performed well in RQ1 on existing single-language bug datasets and tested them on the CLB dataset, analyzing their performance. 
Since our CLB dataset solely consists of Python and Java code, we needed a single-language bug dataset that involves both languages to minimize the impact of PL choice on the experimental results. In the field of software bug research, most datasets focus on C/C++ code, while datasets containing both Python and Java are relatively rare. After reviewing and comparing available options, we ultimately chose CodeNet and CVEfixes as the datasets for our experiments. These are widely used and well-established single-language bug datasets that contain source code in both Java and Python, thereby aligning well with the requirements of our experimental design.

\textbf{CodeNet}~\cite{puri2021codenet}: The CodeNet dataset spans over 50 PLs, including widely used PLs such as C, C++, Java, and Python. It is designed to advance research in code analysis, program understanding, code generation, and bug detection. As one of the largest publicly available code datasets, CodeNet offers extensive applications across various research areas.

\textbf{CVEfixes}~\cite{bhandari2021cvefixes}: The CVEfixes dataset is specifically curated for analyzing and researching software vulnerabilities—considered high-severity software bugs—and their fixes. This dataset encompasses multiple well-known open-source projects, including popular operating systems (e.g., the Linux kernel), server software (e.g., Apache HTTP Server), and PL libraries (e.g., Python’s standard library). The dataset includes code in multiple PLs, such as C, C++, Python, and Java.

Tables \ref{table:codenet_detail} and \ref{table:cvefixes_detail} provide the specific details of Java and Python code in the CodeNet and CVEfixes datasets, respectively. The following are the methods we used to process these two datasets. Since our CLB dataset is balanced (with equal numbers of bug and clean code samples), we processed the CVEfixes and CodeNet datasets to ensure fair experimental results. For the CVEfixes dataset, we used all bug samples. As there are more clean samples than bug samples, we randomly selected an equal number of clean samples to match the bug count, resulting in 3,102 pairs of Java bug and clean samples, and 2,775 pairs of Python bug and clean samples. 
Given the large size of the CodeNet dataset, which contains over 3 million Java and Python code samples, utilizing the entire dataset for fine-tuning would be computationally expensive and impractical. To ensure consistency with our CLB dataset, which contains 11,126 samples, we constructed a subset of the CodeNet dataset for our fine-tuning experiments. Specifically, we first balanced the dataset by randomly selecting an equal number of clean samples to match the number of bug samples. Subsequently, we combined Java and Python samples by class (i.e., merging Java bug samples with Python bug samples, and Java clean samples with Python clean samples), and then randomly selected 10,000 samples each from the bug and clean data to create the final dataset for fine-tuning.

\begin{table}[ht]
\centering
\renewcommand\arraystretch{1.2}
\caption{Details of the CodeNet}
\begin{tabular}{llrr}
\hline
\textbf{Language}                & \textbf{Category} & \textbf{Count*}    & \textbf{Sum}                      \\ \hline
\multirow{2}{*}{Java}   & Buggy      & 305,759  & \multirow{2}{*}{624,479}  \\
                        & Clean   & 318,720  &                          \\ \hline
\multirow{2}{*}{Python} & Buggy      & 1,387,527 & \multirow{2}{*}{2,959,324} \\
                        & Clean   & 1,571,797 &                          \\ \hline
\end{tabular}
    \begin{minipage}{\linewidth}
        \centering
        \vspace{0.2cm}
        {\footnotesize * Only count the results after deduplication.}
    \end{minipage}
\label{table:codenet_detail}
\end{table}

\begin{table}[ht]
\centering
\renewcommand\arraystretch{1.2}
\caption{Details of the CVEfixes (the part of Python and Java)}
\begin{tabular}{llrl}
\hline
\textbf{Language}                & \textbf{Category} & \multicolumn{1}{l}{\textbf{Count}} & \textbf{Sum}                   \\ \hline
\multirow{2}{*}{Java}   & Buggy      & 3,102                     & \multirow{2}{*}{8,387} \\
                        & Clean   & 5,285                     &                       \\ \hline
\multirow{2}{*}{Python} & Buggy      & 2,775                     & \multirow{2}{*}{7,568} \\
                        & Clean   & 4,793                     &                       \\ \hline
\end{tabular}
\label{table:cvefixes_detail}
\end{table}

Specifically, we first randomly split each dataset into training, validation, and test sets with a ratio of 8:1:1. We then selected five CodeLMs that achieved the best performance on the CLB detection task in RQ1 as the subjects of this study.
Next, we fine-tuned these five CodeLMs on the training sets of CVEfixes and CodeNet, respectively, using the corresponding validation sets for model selection and hyperparameter tuning. After fine-tuning, the resulting CodeLMs were directly evaluated on the CLB test set to analyze the extent to which single-language bug knowledge can be transferred to CLB detection. In addition, we used the results obtained by directly fine-tuning the same five CodeLMs on the CLB dataset and evaluating them on the CLB test set as a baseline. This baseline setting is consistent with the experimental setup in RQ1 and serves as a reference for comparing the impact of different fine-tuning datasets on CLB detection performance.

\subsubsection{RQ3: Experimental Setup for Data Scale and Token Length Analysis}
In RQ3, we analyzed the impact of data scale and input sequence length on model performance in the CLB detection task. Consistent with RQ2, we selected the same five CodeLMs that achieved the best performance in RQ1 as the models under study.

\textbf{Data scale analysis.}
Based on the full CLB dataset, we constructed five subsets that contain 20 percent, 40 percent, 60 percent, 80 percent, and 100 percent of the original dataset. Each subset is randomly sampled from the full CLB dataset, while maintaining a similar dataset structure to the original dataset, such as language combinations and label distribution, with a balanced 1:1 ratio between positive and negative samples in each subset. For each subset, the data are randomly split into training, validation, and test sets with a ratio of 8:1:1. For example, when using 60 percent of the full CLB dataset, the subset contains 6,676 code samples, with 5,340 samples for training, 668 samples for validation, and 668 samples for testing. When using the full dataset, all 11,126 code samples are included, with 8,902 samples for training, 1,112 samples for validation, and 1,112 samples for testing. We fine-tuned the models on the corresponding training sets and evaluated their performance on the test sets to analyze how model performance changes under different data scale settings.

\textbf{Token sequence length analysis.}
For the input length analysis, we used the full CLB dataset and keep the same splits for the training, validation, and test sets. During model training, validation, and testing, we limited the maximum input token length to 128, 256, 384, and 512. When the input sequence length exceeds the predefined limit, a consistent truncation strategy is applied, where only tokens within the maximum length are kept and the remaining tokens are discarded. Through this experimental setup, we analyze the impact of different input length constraints on CLB detection performance.

\subsubsection{RQ4: Experimental Setup for Evaluating the Effect of Code Comments}
The experimental setup for RQ4 is consistent with that of RQ1, with the only difference being the dataset preprocessing strategy. In this RQ, we used the original code dataset without removing code comments for CodeLMs fine-tuning and testing, in order to evaluate the impact of code comment information on CLB detection performance.

\section{Study Results}\label{chap:results}

\subsection{Overview of the Obtained Dataset}\label{OverviewOfDataSet}
Based on the CLB sample mining method described in Section \ref{data_collection}, we constructed a dataset for CLB detection. The dataset contains a total of 5,563 function pairs, comprising 11,126 function instances. Each pair consists of a real-world buggy function and its corresponding fixed version (clean function), forming a complete bug–fix pair that can be directly used for supervised learning and model evaluation.

\begin{figure}[ht]
    \centering
    \begin{minipage}[b]{0.24\textwidth}
        \centering
        \includegraphics[width=\linewidth]{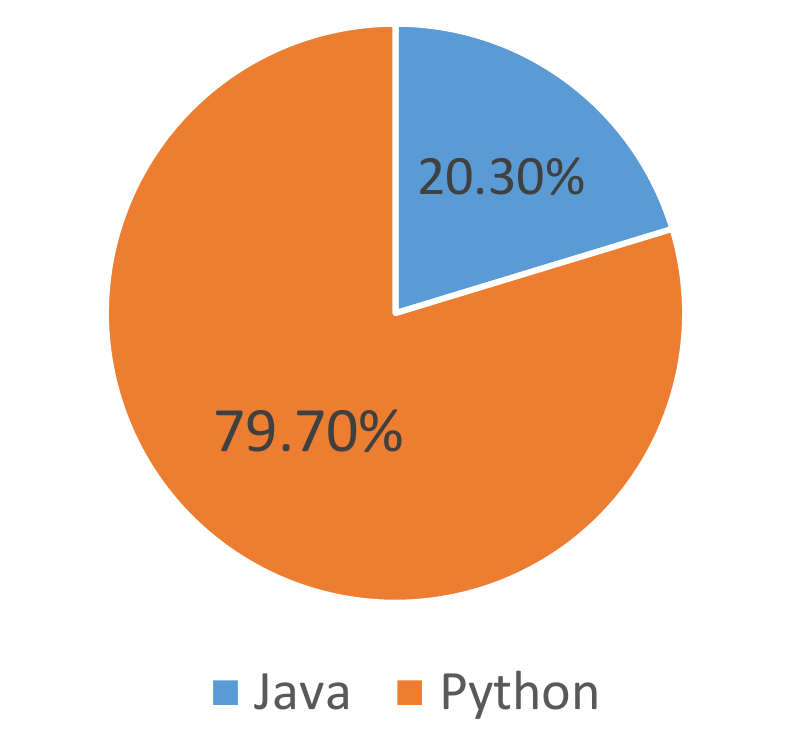}
        \subcaption{Proportion of Java and Python samples} \label{fig:sub1}
    \end{minipage}
    \hfill
    \begin{minipage}[b]{0.24\textwidth}
        \centering
        \includegraphics[width=\linewidth]{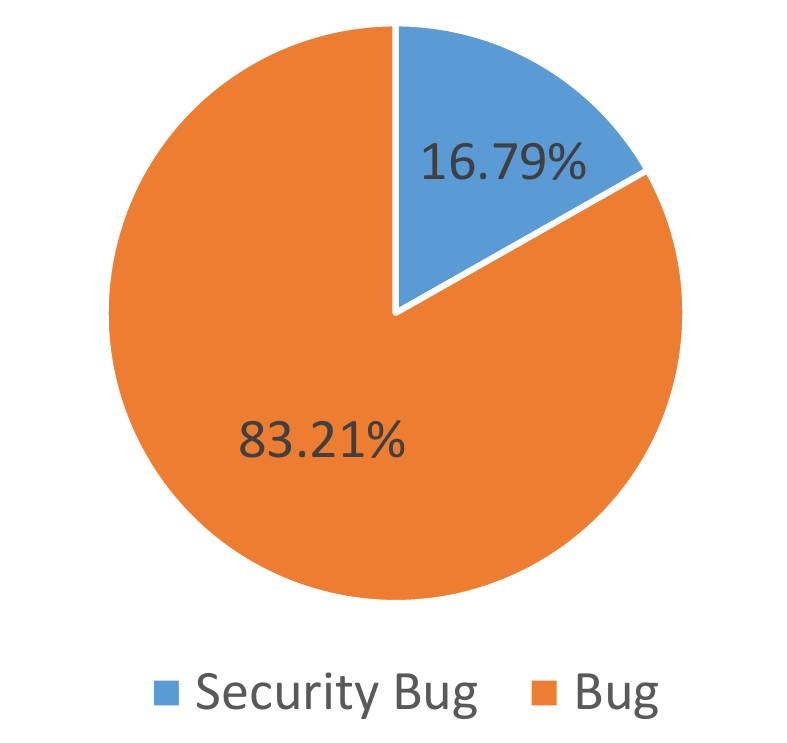}
        \subcaption{Proportion of Bug and Security Bug} \label{fig:sub2}
    \end{minipage}
    \hfill
    \begin{minipage}[b]{0.24\textwidth}
        \centering
        \includegraphics[width=\linewidth]{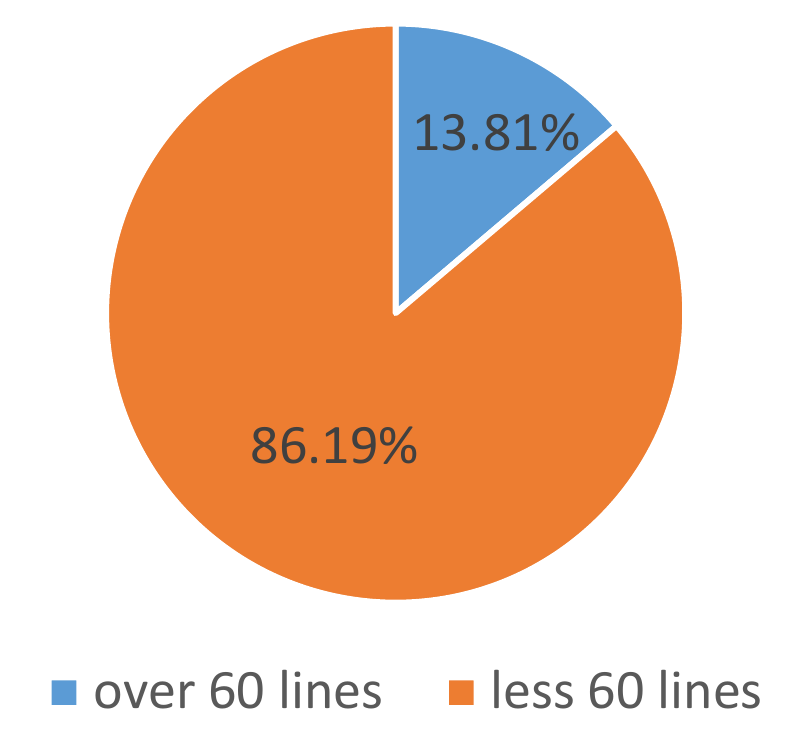}
        \subcaption{Length distribution of buggy code} \label{fig:sub3}
    \end{minipage}
    \hfill
    \begin{minipage}[b]{0.24\textwidth}
        \centering
        \includegraphics[width=\linewidth]{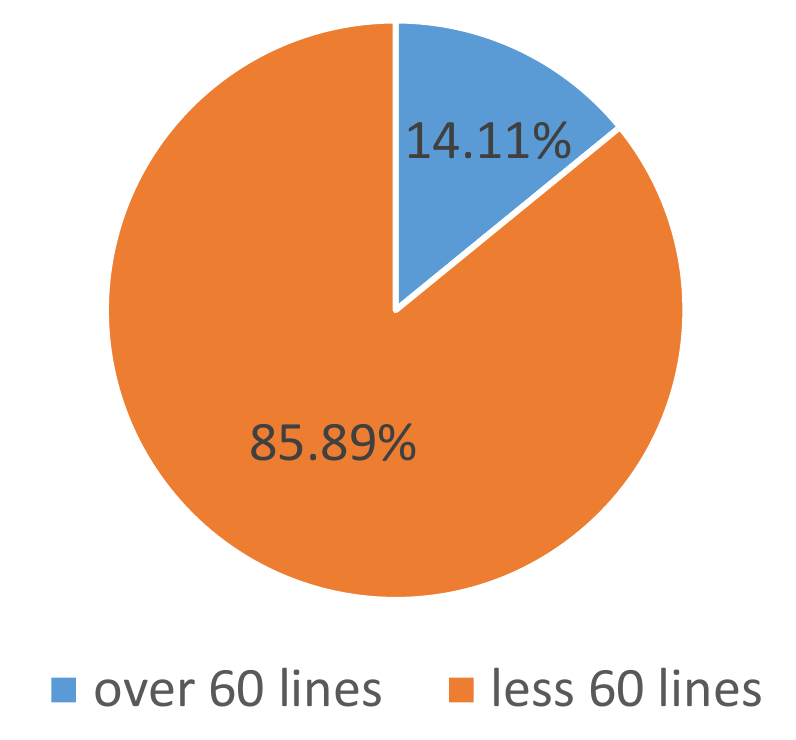}
        \subcaption{Length distribution of clean code} \label{fig:sub4}
    \end{minipage}
    \caption{Details of our dataset}
    \label{fig:pie_charts}
\end{figure}

In terms of PL composition, the dataset primarily includes two mainstream PLs: Python and Java. Specifically, there are 4,464 Python function pairs (79.70\%) and 1,137 Java function pairs (20.30\%), as illustrated in Figure \ref{fig:pie_charts}(a). Since Python and Java often interact with C/C++ code via precompiled libraries - such as \texttt{.so} files on Linux and \texttt{.dll} files on Windows - rather than directly including C/C++ source code, we limit our analysis to instances where Python and Java invoke C/C++ libraries, as well as cases where Java invokes Python, in order to streamline tool implementation and enhance efficiency. All data samples were collected from open-source projects on GitHub, covering a total of 190 real-world repositories across various application domains and coding styles. Among them, 163 projects contributed to Python code, while 28 projects contributed to Java code. Some projects, such as \texttt{RLbot}, contain code in both languages.

In addition to general bugs, our dataset encompasses samples specifically related to security vulnerabilities. A total of 934 function pairs are associated with high-severity issues, primarily involving security vulnerability fixes. These samples originate from 233 security-related commits and are of significant practical value. These data serve as a foundation for research on cross-language security vulnerability detection and analysis. These data suggest that our data collection methodology is applicable to the security domain and has the potential to facilitate continuous and scalable collection of security-related samples.

\begin{figure}[ht]
    \centering
    \includegraphics[width=0.6\linewidth]{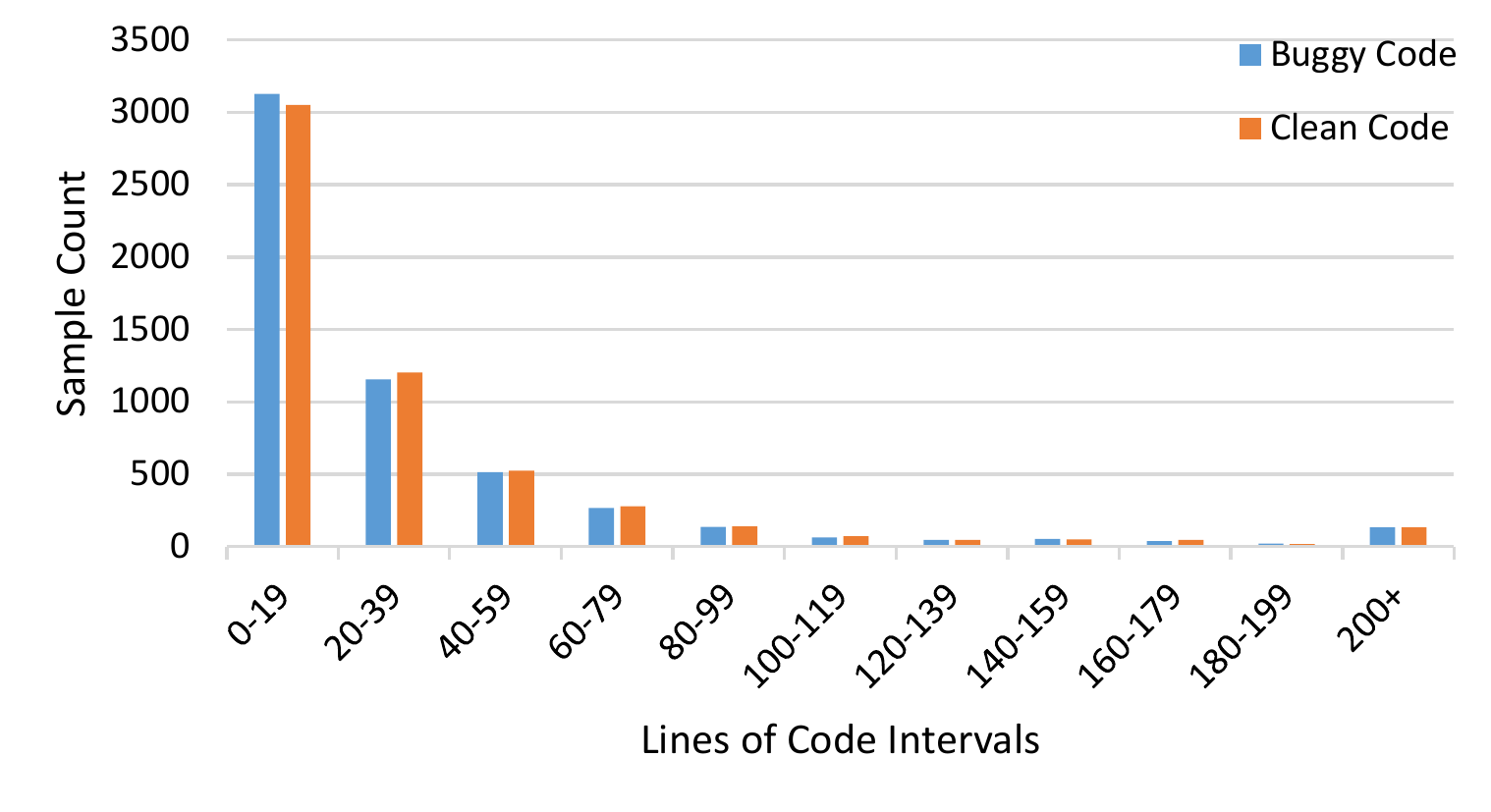}
    \caption{The line count distribution of the CLB dataset with the unit of measurement being lines}
    \label{fig:size_line}
\end{figure}

\begin{figure*}[ht]
    \centering
    \includegraphics[width=1\linewidth]{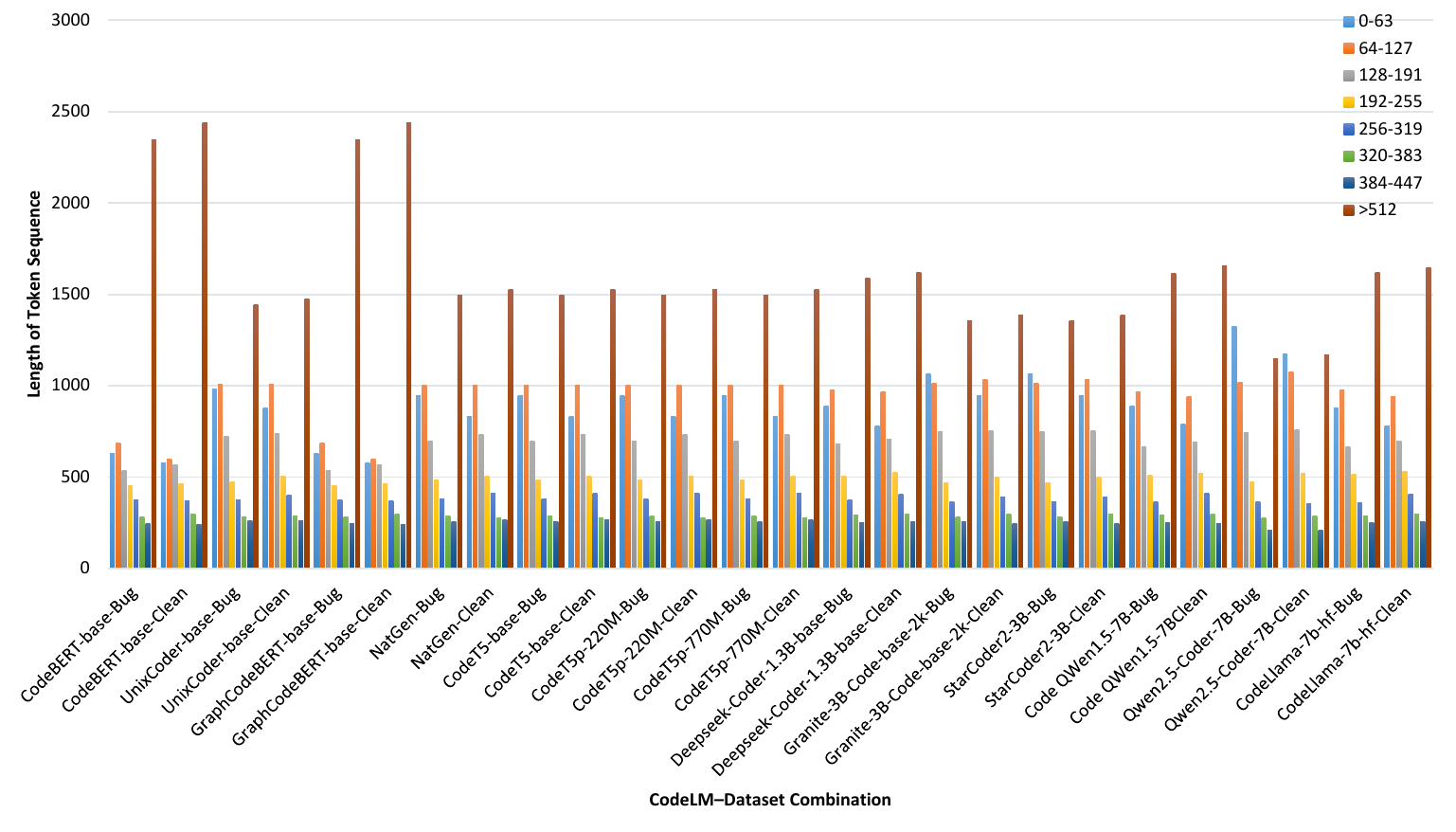}
    \caption{The token sequence length distribution of tokenization results produced by different models' tokenizers on the data within the CLB dataset}
    \label{fig:token_distribution}
\end{figure*}

To further characterize the dataset, we conducted statistical analysis on code length and token sequence length distribution. Figure~\ref{fig:size_line} shows the distribution of code length (in lines) for both buggy and clean functions. After removing blank lines and comments, 86.19\% of buggy functions and 85.89\% of clean functions contain fewer than 60 lines, making them suitable for processing by lightweight models. We chose 60 lines as the cutoff between short and long functions because the distribution shows a clear drop in frequency around this point, marking a natural inflection. This indicates that short to medium-length functions are more common, while functions longer than 60 lines are way fewer. Thus, 60 lines serve as a statistically meaningful and practical boundary to distinguish general functions from longer ones. Meanwhile, the dataset also includes a number of complex samples: 133 buggy functions (2.39\%) and 132 clean functions (2.37\%) exceed 200 lines, providing challenging cases for evaluating model performance on long code sequences.

Figure~\ref{fig:token_distribution} presents the tokenized sequence length distribution using 13 widely adopted pre-trained model tokenizers. The results indicate that most samples contain fewer than 512 tokens after tokenization, aligning well with the input length limitations of mainstream Transformer-based architectures. Although different tokenizers produce slightly varied results, the overall distribution trends are consistent, supporting comparative evaluations across multiple models. It is important to note that all statistics shown in Figures \ref{fig:size_line} and \ref{fig:token_distribution} are based on code with blank lines and comments removed.

While the above analyses characterize the CLB dataset from a structural and statistical perspective, they do not fully reflect the semantic nature of CLBs themselves. To provide a more comprehensive overview of the obtained dataset, we further analyze the observable symptoms and underlying root causes of CLBs.

\begin{figure}[ht]
    \centering
    \includegraphics[width=0.9\linewidth]{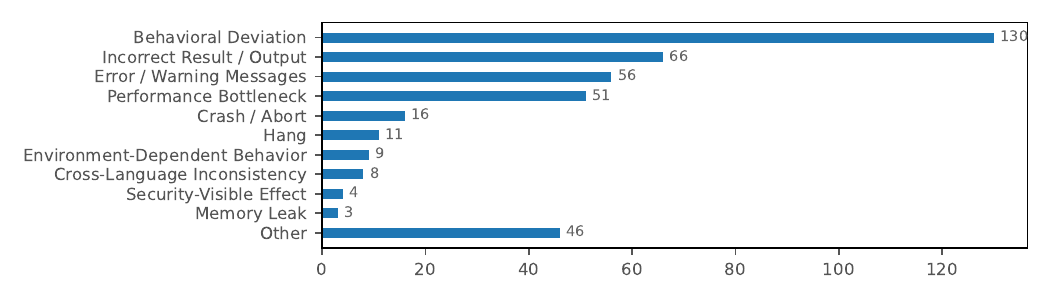}
    \caption{Distribution of Cross-Language Bug Symptoms in the CLB Dataset}
    \label{fig:sym}
\end{figure}

\begin{figure}[ht]
    \centering
    \includegraphics[width=0.9\linewidth]{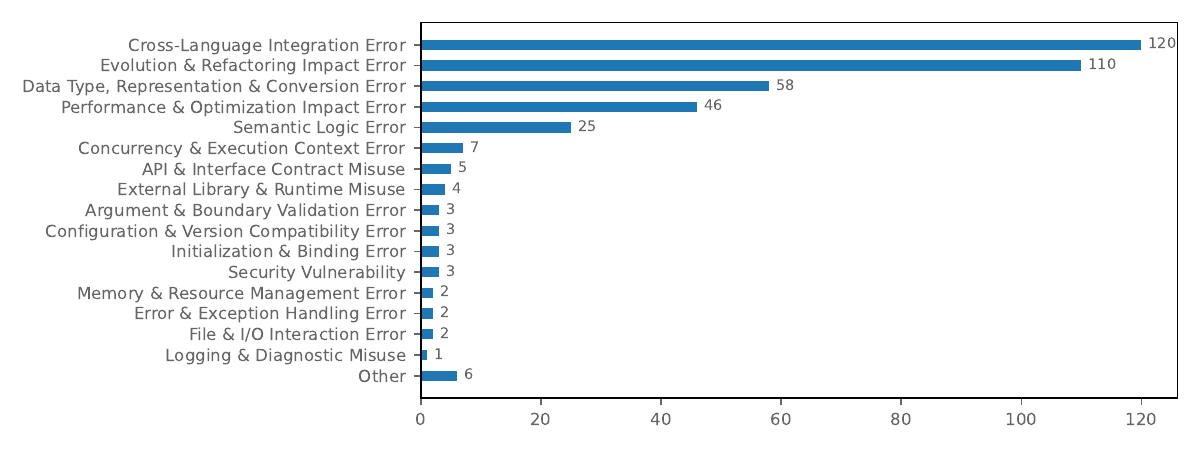}
    \caption{Distribution of Root Causes of CLBs in the CLB Dataset}
    \label{fig:root-cause}
\end{figure}

In addition, we conduct a systematic analysis of the symptoms and root causes of CLBs in the CLB dataset, following the methodology proposed by Yang et al \cite{yang2025dissecting}. Since manually inspecting all bug instances is time-consuming, we randomly sample 400 bug cases from the entire dataset for analysis. This sample size is statistically significant with a 95\% confidence level and a 5\% margin of error. All sampled bug cases are manually examined and categorized by the first and second authors, and any disagreements are resolved through discussion until a consensus is reached. During symptom and root cause classification, we adopt the taxonomy from Yang et al. and make minor adjustments to several categories to better fit the characteristics of the CLB dataset. The detailed distributions of symptoms and root causes are illustrated in Figures \ref{fig:sym} and \ref{fig:root-cause}, respectively.

Overall, CLBs exhibit relatively concentrated distributions in both their root causes and observable symptoms, while also demonstrating clear characteristics of cross-language interactions. These distributions offer an intuitive view of how CLBs are triggered and how they manifest at runtime.

Regarding observable symptoms, CLBs tend to manifest as non-fatal yet hard-to-detect runtime anomalies. Behavioral deviation is the most frequently observed symptom, indicating that many CLBs do not lead to crashes or explicit error reports, but instead appear as unexpected changes in program behavior or output. Error or warning messages and incorrect results or outputs are also commonly observed, showing that some CLBs are exposed through explicit but non-terminating signals. In contrast, overt failures such as crashes and memory leaks occur in relatively fewer cases, suggesting that CLBs are often more hidden and therefore more challenging to debug and localize. Additionally, it should be noted that the ``Other” category accounts for approximately 10\% of the sampled instances. This category includes various uncommon bug symptoms that cannot be classified into the existing major categories, such as (but not limited to) logging and observability issues, API deprecation mismatches at language boundaries, and metric reporting failures. These types are grouped into “Other” mainly due to their very small sample sizes. Furthermore, this category also includes cases where the specific symptom cannot be determined due to insufficient descriptions in the original commit messages and issue reports.

From the perspective of root causes, issues directly related to cross-language integration dominate the dataset. Among them, cross-language integration errors constitute the largest category, indicating that many bugs originate from inconsistencies at language boundaries, such as cross-language interface invocations and misalignment of data and control flows. In addition, errors introduced by system evolution and refactoring also account for a substantial number of cases, suggesting that implicit cross-language assumptions are easily violated during the long-term evolution of multilingual systems. Meanwhile, data type and cross-language representation errors, as well as performance and optimization related errors, are also observed in non-negligible quantities, reflecting the complexity of data representation and execution semantics in cross-language systems.

The statistical analysis of CLB symptoms and root causes indicate that bugs in the CLB dataset exhibit strong cross-language characteristics. Their root causes are mainly concentrated at language boundaries, system evolution, and data representation, while their symptoms are more likely to appear as behavioral deviations rather than immediate failures. This analysis further highlights the inherent complexity of cross-language bugs in real-world software systems and provides empirical support for the necessity of dedicated modeling and research on CLBs.

In summary, this dataset demonstrates significant strengths in terms of size, quality, and language diversity. The bug-fix pairing structure naturally fits various supervised learning tasks and supports a wide range of research objectives, such as bug detection, localization, and repair.

\subsection{RQ1: CodeLM Performance after Fine-Tuning}\label{RQ1Results}

\subsubsection{Performance Metrics of CodeLMs}

We fine-tuned 13 well-known open-source CodeLMs and evaluated their performance in detecting CLBs.
In addition, we include GPT-4o-mini as an inference-only baseline, which is evaluated on the same test set using a prompt-based reasoning strategy, without any task-specific fine-tuning. The detailed results are presented in Table \ref{table:rq1}, note that the entries for GPT-4o-mini only report inference performance.
The detailed comparisons of the fine-tuned CodeLMs performance metrics, alongside the GPT-4o-mini baseline, are presented in the following.

\begin{table*}[ht]
\centering
\renewcommand\arraystretch{1.2}
\caption{Performance of CodeLMs in detecting CLBs after fine-tuning on the CLB dataset without code comments}
{
\begin{tabular}{lccccc}
\hline
\textbf{CodeLM} & \textbf{Acc} & \textbf{Prec} & \textbf{Rec} & \textbf{F1} & \textbf{AUC} \\ \hline
GPT-4o-mini (Baseline) & \grayA{0.5196} & \grayA{0.5190} & \grayA{0.5357} & \grayA{0.5272} & \grayA{0.5196} \\
CodeBERT-base & \grayA{0.6331} & \grayA{0.6022} & \grayA{0.7842} & \grayA{0.6813} & \grayA{0.7131} \\
GraphCodeBERT-base & \grayA{0.6517} & \grayA{0.6128} & \grayA{0.8241} & \grayA{0.7029} & \grayA{0.7447} \\
UniXcoder-base & \grayA{0.7172} & \grayA{0.6839} & \grayA{0.8079} & \grayA{0.7407} & \grayA{0.8088} \\
CodeT5-base & \grayA{0.6454} & \grayA{0.6350} & \grayA{0.6840} & \grayA{0.6586} & \grayA{0.7238} \\
CodeT5p-220M & \grayA{0.6364} & \grayA{0.6148} & \grayA{0.7307} & \grayA{0.6678} & \grayA{0.7313} \\
CodeT5p-770M & \grayA{0.5302} & \grayA{0.5238} & \grayA{0.8779} & \grayA{0.6561} & \grayA{0.5432} \\
NatGen & \grayA{0.6364} & \grayA{0.6095} & \grayA{0.7594} & \grayA{0.6763} & \grayA{0.7278} \\
Deepseek-Coder-1.3B-base & \grayA{0.5498} & \grayA{0.5255} & \grayA{0.7063} & \grayA{0.6026} & \grayA{0.6125} \\
Granite-3B-Code-base-2k & \grayA{0.5516} & \grayA{0.5275} & \grayA{0.6951} & \grayA{0.5998} & \grayA{0.5865} \\
StarCoder2-3B & \grayA{0.5561} & \grayA{0.5358} & \grayA{0.7124} & \grayA{0.6116} & \grayA{0.5889} \\
CodeQWen1.5-7B & \grayA{0.5229} & \grayA{0.5043} & \grayA{0.7580} & \grayA{0.6057} & \grayA{0.5706} \\
Qwen2.5\_Coder-7B & \grayA{0.5238} & \grayA{0.5066} & \grayA{0.5669} & \grayA{0.5350} & \grayA{0.5676} \\
CodeLlama-7b-hf & \grayA{0.4979} & \grayA{0.5079} & \grayA{0.5269} & \grayA{0.5172} & \grayA{0.5442} \\ \hline
\end{tabular}
}
\label{table:rq1}
\end{table*}

\textbf{Accuracy}: As shown in Table \ref{table:rq1}, UniXcoder-base achieves the highest accuracy of 0.7172, followed by GraphCodeBERT-base at 0.6517. In contrast, CodeLlama-7b-hf yields the lowest accuracy among the fine-tuned CodeLMs at 0.4979. For reference, the inference-only GPT-4o-mini baseline achieves an accuracy of 0.5196, indicating that fine-tuned CodeLMs effectively enhance the overall prediction accuracy and significantly outperform the general LLM under zero-shot settings in the CLB detection task.


\textbf{Precision}: The precision of the fine-tuned models varies, with UniXcoder-base achieving the highest precision at 0.6839, followed by CodeT5-base at 0.6350 and CodeT5p-220M at 0.6148. These results suggest that fine-tuning enables the models to effectively identify positive instances while controlling false positives. In comparison, GPT-4o-mini achieves a precision of 0.5190, which is surpassed by almost all fine-tuned CodeLMs.


\textbf{Recall}: Recall reflects the model's ability to capture positive instances. Several fine-tuned CodeLMs demonstrate strong recall, such as CodeT5p-770M (0.8779) and GraphCodeBERT-base (0.8241), indicating high sensitivity in recognizing CLBs. Conversely, Qwen2.5\_Coder-7B and CodeLlama-7b-hf show lower recall scores of 0.5669 and 0.5269, respectively. The GPT-4o-mini baseline achieves a recall of 0.5357, further highlighting that task-specific adaptation generally leads to higher sensitivity in bug detection.

\textbf{F1 score}: 
UniXcoder-base achieves the highest F1 score of 0.7407, demonstrating an excellent balance between precision and recall. GraphCodeBERT-base also performs well with an F1 score of 0.7029.
In comparison, the F1 score of GPT-4o-mini is 0.5272, which is consistently lower than those of almost all fine-tuned CodeLMs, highlighting the effectiveness of task-specific fine-tuning for CLB detection.

\textbf{AUC}: The AUC metric further confirms the discriminative capability of the fine-tuned CodeLMs. UniXcoder-base reaches the highest AUC of 0.8088, indicating a strong ability to distinguish between positive and negative samples, particularly in challenging scenarios such as semantically similar code snippets. Similarly, the AUC of GPT-4o-mini is 0.5196, which is notably lower than those of the fine-tuned CodeLMs, indicating limited discriminative capability under inference-only settings.

\subsubsection{Performance of Small and Large CodeLMs}
As shown in the results of Table \ref{table:rq1}, the impact of fine-tuning varies across different CodeLMs. We further explored the differences in fine-tuning effects based on the CodeLM's size and complexity. According to the definition in~\cite{zhang2024comprehensive}, models with fewer than 1B parameters are categorized as small LMs, while models with 1B parameters or more are considered as large LMs.

\textbf{Small CodeLMs}: Small-scale CodeLMs, such as UniXcoder-base, GraphCodeBERT-base, CodeBERT-base, CodeT5p-220M, and NatGen, exhibited strong overall performance. In particular, UniXcoder-base demonstrated the best results across almost all metrics, suggesting that small CodeLMs can gain more knowledge transfer and performance enhancement from fine-tuning on the CLB dataset. On the other hand, among the small CodeLMs, CodeT5p-770M, with 770M parameters, showed the poorest overall performance. Despite having the best recall metric among all models (0.8779), it performed worse in the other four metrics. The result suggests that, for the CLB detection task, large CodeLMs do not necessarily outperform small CodeLMs in terms of overall performance.

\textbf{Large CodeLMs}: Among the large CodeLMs, those with 3B or fewer parameters, specifically Deepseek-Coder-1.3B-base, Granite-3B-Code-base-2k, and StarCoder2-3B, showed moderate performance, yet still outperformed the baseline model GPT-4o-mini on the CLB detection task. As for the remaining three 7B large models — CodeQWen1.5-7B, Qwen2.5\_Coder-7B, and CodeLlama-7b-hf — their performance was relatively limited compared to the smaller models, only slightly surpassing GPT-4o-mini on certain metrics while falling behind it on others.

\begin{tcolorbox}
\textbf{Finding 1}:
Among the 13 CodeLMs, UniXcoder-base and GraphCodeBERT-base demonstrate strong performance, with UniXcoder-base achieving the best overall results across all metrics. The performance of the 7B models, such as Qwen2.5\_Coder-7B and CodeLlama-7b-hf are relatively limited. Overall, within the scope of our experimental setup, small CodeLMs tend to achieve better performance than large ones in the CLB detection task.
The inference-only GPT-4o-mini baseline is consistently outperformed by fine-tuned CodeLMs, highlighting the importance of task-specific adaptation.

\end{tcolorbox}

\subsection{RQ2: Impact of Different Fine-Tuning Datasets on CLB Detection Performance}\label{RQ2Results}

To answer RQ2, we further investigate whether CodeLMs fine-tuned on single-language bug datasets can effectively improve their ability to detect CLB. Specifically, five CodeLMs that perform well in RQ1 are selected. These CodeLMs are fine-tuned on 2 representative single-language bug datasets, namely CodeNet and CVEfixes, as well as on the CLB dataset constructed in this work. After fine-tuning, all CodeLMs are evaluated on the CLB test set to ensure a fair comparison of different fine-tuning strategies in the CLB detection task.

Tables \ref{table:codenet} and \ref{table:cvefixes} report the experimental results. When the CodeLMs are fine-tuned only on single-language bug datasets, either CodeNet or CVEfixes, their detection performance on the CLB dataset is generally limited. In particular, the accuracy and precision of most CodeLMs are close to 0.5, and the AUC values also remain around 0.5. This indicates that the CodeLMs achieve near random performance in the CLB detection task. Although some CodeLMs achieve relatively higher recall values and slightly improved F1 scores, this improvement is limited and is not reflected in the AUC metric. This result suggests that the CodeLMs fail to achieve an effective balance between precision and recall in CLB detection.

\begin{table*}[ht]
\centering
\renewcommand\arraystretch{1.2}
\caption{CLB Detection Performance of CodeLMs Fine-Tuned on CodeNet and the CLB Dataset}
\resizebox{\textwidth}{!}{ 
\begin{tabular}{lcccccccccc}
\hline
\multicolumn{1}{c}{\multirow{2}{*}{\textbf{CodeLM}}} & \multicolumn{5}{c}{\textbf{Fine-tuned in CodeNet}} & \multicolumn{5}{c}{\textbf{Fine-tuned in CLB Dataset}} \\ \cline{2-11}
\multicolumn{1}{c}{} & \textbf{Acc} & \textbf{Prec} & \textbf{Rec} & \textbf{F1} & \textbf{AUC} & \textbf{Acc} & \textbf{Prec} & \textbf{Rec} & \textbf{F1} & \textbf{AUC} \\ \hline
CodeBERT-base      & \grayA{0.5009} & \grayA{0.5006} & \grayA{0.7253} & \grayA{0.5924} & \grayA{0.4869} & \grayA{0.6331} & \grayA{0.6022} & \grayA{0.7842} & \grayA{0.6813} & \grayA{0.7131} \\
GraphCodeBERT-base & \grayA{0.5108} & \grayA{0.5081} & \grayA{0.6732} & \grayA{0.5792} & \grayA{0.5012} & \grayA{0.6517} & \grayA{0.6128} & \grayA{0.8241} & \grayA{0.7029} & \grayA{0.7447} \\
UniXcoder-base     & \grayA{0.4973} & \grayA{0.4982} & \grayA{0.7343} & \grayA{0.5936} & \grayA{0.5028} & \grayA{0.7172} & \grayA{0.6839} & \grayA{0.8079} & \grayA{0.7407} & \grayA{0.8088} \\
CodeT5p-220M       & \grayA{0.5162} & \grayA{0.5105} & \grayA{0.7864} & \grayA{0.6191} & \grayA{0.5072} & \grayA{0.6364} & \grayA{0.6148} & \grayA{0.7307} & \grayA{0.6678} & \grayA{0.7313} \\
NatGen             & \grayA{0.4982} & \grayA{0.4987} & \grayA{0.6934} & \grayA{0.5802} & \grayA{0.5010} & \grayA{0.6364} & \grayA{0.6095} & \grayA{0.7594} & \grayA{0.6763} & \grayA{0.7278} \\
\hline
\end{tabular}
}
\label{table:codenet}
\end{table*}

\begin{table*}[ht]
\centering
\renewcommand\arraystretch{1.2}
\caption{CLB Detection Performance of CodeLMs Fine-Tuned on CVEfixes and the CLB Dataset}
\resizebox{\textwidth}{!}{ 
{
\begin{tabular}{lcccccccccc}
\hline
\multicolumn{1}{c}{\multirow{2}{*}{\textbf{CodeLM}}} & \multicolumn{5}{c}{\textbf{Fine-tuned in CVEfixes}} & \multicolumn{5}{c}{\textbf{Fine-tuned in CLB Dataset}} \\ \cline{2-11}
\multicolumn{1}{c}{} & \textbf{Acc} & \textbf{Prec} & \textbf{Rec} & \textbf{F1} & \textbf{AUC} & \textbf{Acc} & \textbf{Prec} & \textbf{Rec} & \textbf{F1} & \textbf{AUC} \\ \hline
CodeBERT-base      & \grayA{0.5036} & \grayA{0.5022} & \grayA{0.8366} & \grayA{0.6276} & \grayA{0.5024} & \grayA{0.6331} & \grayA{0.6022} & \grayA{0.7842} & \grayA{0.6813} & \grayA{0.7131} \\
GraphCodeBERT-base & \grayA{0.4937} & \grayA{0.4959} & \grayA{0.7684} & \grayA{0.6028} & \grayA{0.4999} & \grayA{0.6517} & \grayA{0.6128} & \grayA{0.8241} & \grayA{0.7029} & \grayA{0.7447} \\
UniXcoder-base     & \grayA{0.5018} & \grayA{0.5011} & \grayA{0.7864} & \grayA{0.6122} & \grayA{0.4901} & \grayA{0.7172} & \grayA{0.6839} & \grayA{0.8079} & \grayA{0.7407} & \grayA{0.8088} \\
CodeT5p-220M       & \grayA{0.5018} & \grayA{0.5012} & \grayA{0.7504} & \grayA{0.6010} & \grayA{0.5053} & \grayA{0.6364} & \grayA{0.6148} & \grayA{0.7307} & \grayA{0.6678} & \grayA{0.7313} \\
NatGen             & \grayA{0.5152} & \grayA{0.5094} & \grayA{0.8217} & \grayA{0.6289} & \grayA{0.4938} & \grayA{0.6364} & \grayA{0.6095} & \grayA{0.7594} & \grayA{0.6763} & \grayA{0.7278} \\
\hline
\end{tabular}}}

\label{table:cvefixes}
\end{table*}

In contrast, when the CodeLMs are directly fine-tuned on the CLB dataset, their performance in CLB detection improves significantly. All CodeLMs show consistent improvements in accuracy, precision, F1 score, and AUC. For example, UniXcoder-base achieves an F1 score of 0.7407 and an AUC of 0.8088 after fine-tuning on the CLB dataset, which is clearly better than its performance after fine-tuning on single-language datasets. Similar performance gains are observed for the other CodeLMs, indicating that fine-tuning on CLB data provides a stable and effective improvement for CLB detection.

Overall, these results indicate that fine-tuning only on single-language bug datasets does not transfer well to the CLB detection task. CLB usually arise from interactions and dependencies between different PLs, and their characteristics are fundamentally different from those of single-language bugs. Therefore, CodeLMs can effectively learn relevant patterns and achieve reliable detection when they are fine-tuned on CLB datasets that explicitly contain cross-language semantics and interaction information.

\begin{tcolorbox}

\textbf{Finding 2}: CodeLMs fine-tuned on single-language bug datasets such as CodeNet and CVEfixes show limited performance in the CLB detection task, with overall results close to random performance. In contrast, fine-tuning on the CLB dataset significantly improves detection performance. This finding indicates that CLB detection requires dedicated datasets and training strategies, and single-language bug data alone cannot adequately support this task.
\end{tcolorbox}

\subsection{RQ3: Performance Influencing Factors of Fine-tuned CodeLMs}\label{RQ3Results}

A key limitation faced in this research is the token sequence length constraint in CodeLMs, leading to truncation of longer data samples, which prevents the fine-tuned CodeLMs from fully understanding the code's context. Additionally, due to the vast pre-training data of the CodeLMs and the relatively limited size of our CLB dataset (only 5,563 code pairs) for fine-tuning, we designed experiments to investigate how dataset size and token sequence length impact model performance in CLB detection.

\subsubsection{Impact of the Fine-tuning Dataset Size}

\begin{figure}[ht]
    \centering
    \begin{minipage}[b]{0.49\textwidth}
        \centering
        \includegraphics[width=\linewidth]{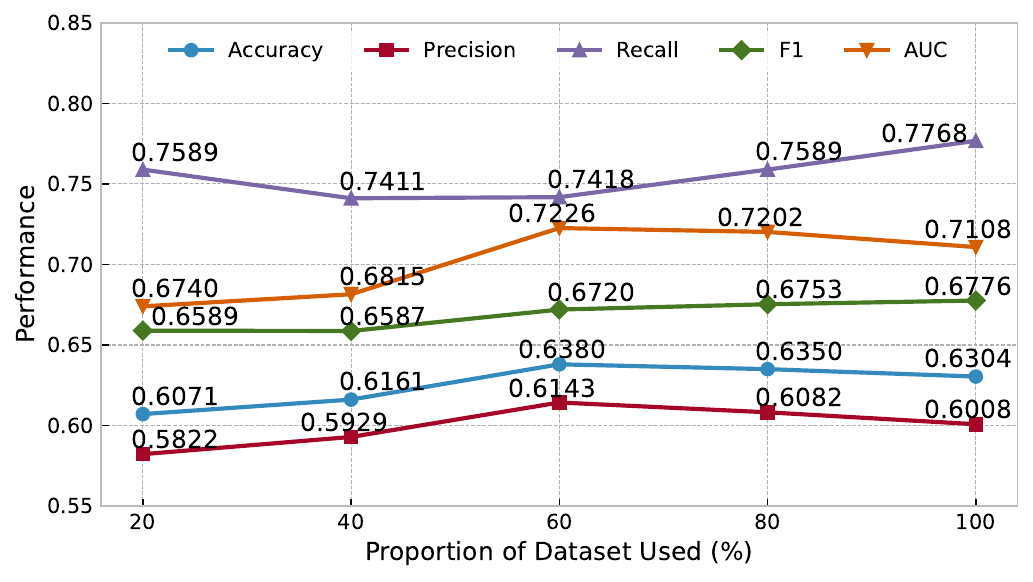}
        \subcaption{CodeBERT-base} \label{fig:sub1}
    \end{minipage}
    \hfill
    \begin{minipage}[b]{0.49\textwidth}
        \centering
        \includegraphics[width=\linewidth]{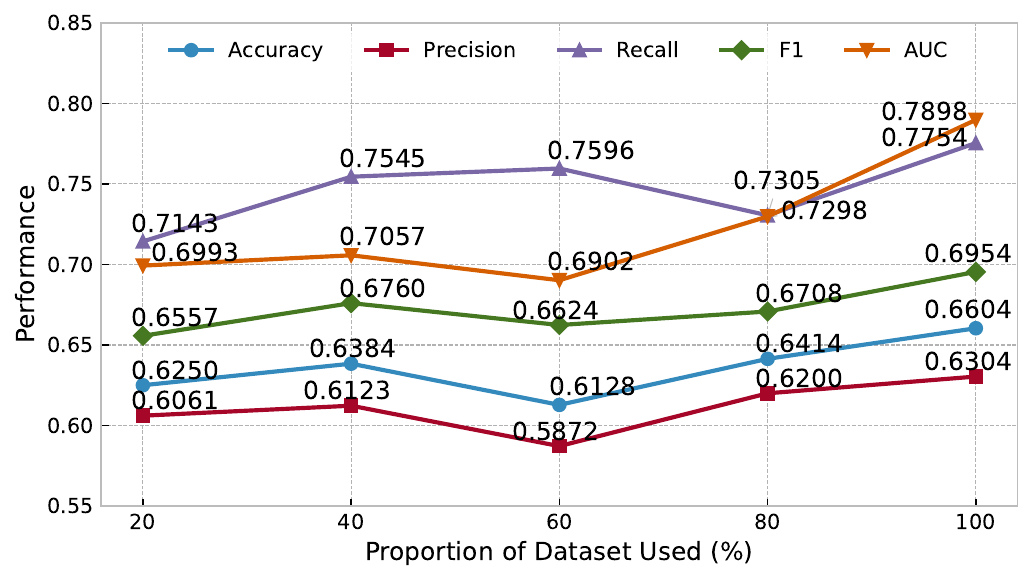}
        \subcaption{GraphCodeBERT-base} \label{fig:sub2}
    \end{minipage}

    \vspace{1em} 

    \begin{minipage}[b]{0.49\textwidth}
        \centering
        \includegraphics[width=\linewidth]{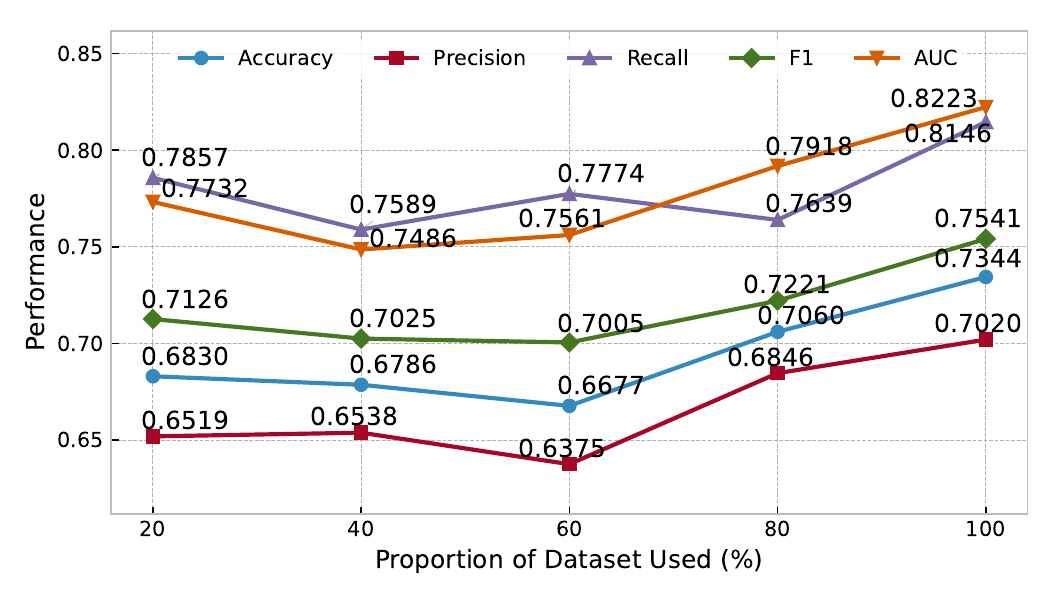}
        \subcaption{UniXcoder-base} \label{fig:sub3}
    \end{minipage}
    \hfill
    \begin{minipage}[b]{0.49\textwidth}
        \centering
        \includegraphics[width=\linewidth]{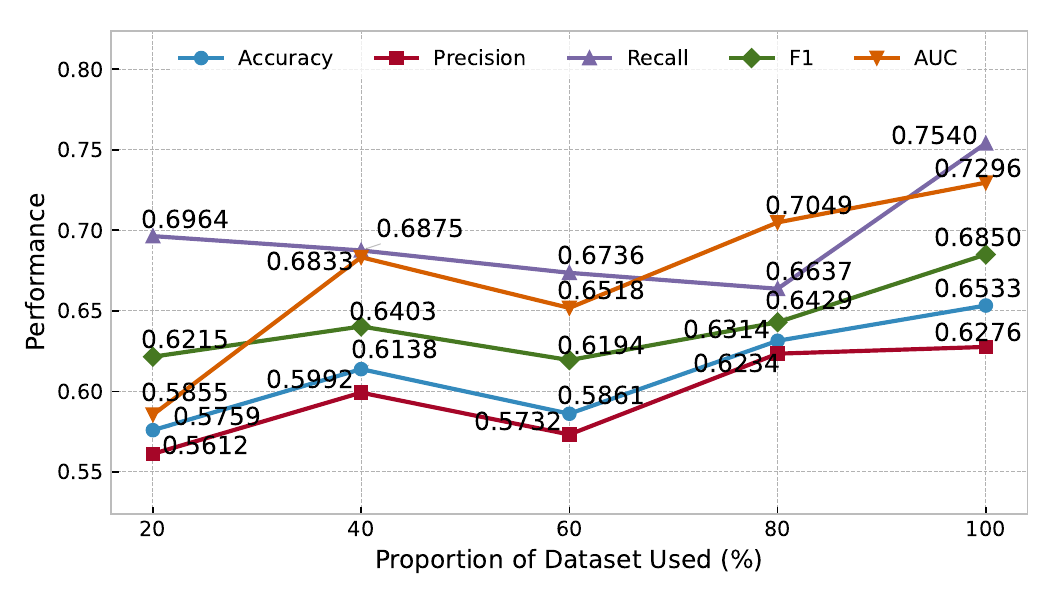}
        \subcaption{CodeT5p-220M} \label{fig:sub4}
    \end{minipage}

    \vspace{1em}

    \begin{minipage}[b]{0.49\textwidth}
        \centering
        \includegraphics[width=\linewidth]{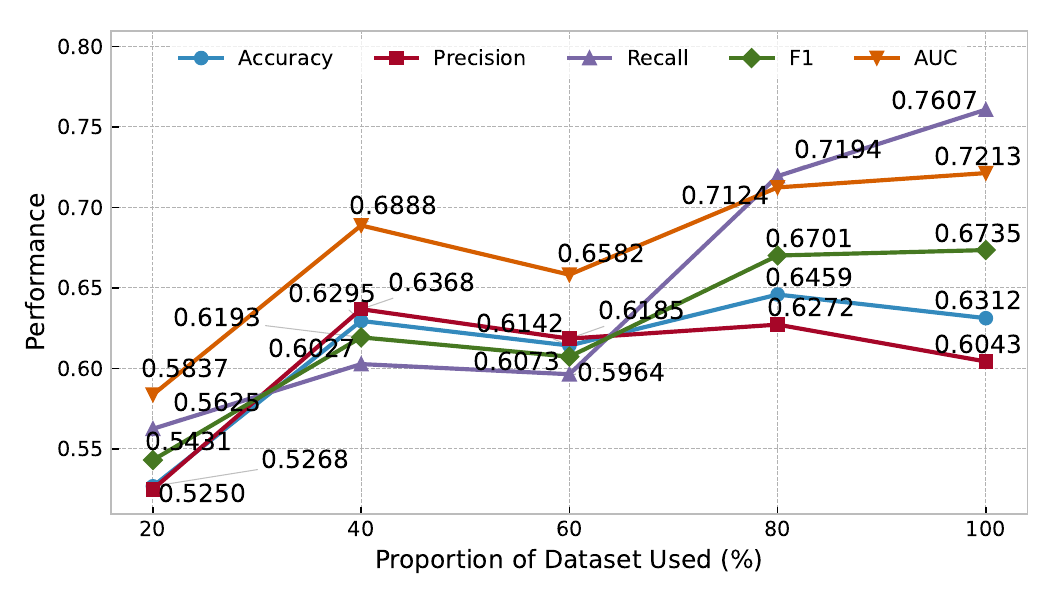}
        \subcaption{NatGen} \label{fig:sub5}
    \end{minipage}

    \caption{Performance of CodeLMs in detecting CLBs with different dataset sizes}
    \label{fig:rq3_datasize}
\end{figure}

This experiment primarily explores the impact of dataset size on the performance of fine-tuned CodeLMs in CLB detection. The CodeLMs were fine-tuned on 20\%, 40\%, 60\%, 80\%, and 100\% of our CLB dataset, respectively. Detailed experimental results are reported in Figure \ref{fig:rq3_datasize}. An in-depth analysis of the experimental results are presented as follows.

\textbf{The overall trends of accuracy, F1, and AUC}: From Figure \ref{fig:rq3_datasize}, we can observe that as the fine-tuning dataset size increases, all CodeLMs show varying degrees of improvement in accuracy, F1, and AUC. Specifically, UniXcoder-base maintains the best performance across all dataset sizes, with its accuracy increasing from 0.6830 to 0.7344, F1 from 0.7126 to 0.7541, and AUC from 0.7732 to 0.8223. This indicates that UniXcoder-base performs well even with a small dataset and continues to improve with more data. In contrast, GraphCodeBERT-base and CodeBERT-base show relatively smaller improvements. CodeBERT-base, in particular, only increases its accuracy by 0.0233 (0.6304-0.6071) and F1 by 0.0187 (0.6776-0.6589), suggesting that it already has strong generalization ability with a small dataset, and the benefits of additional data are limited. Besides, CodeT5p-220M and NatGen exhibit lower performance with small dataset (20\%), but as the dataset size expands to the size of the full dataset, their accuracy, F1, and AUC improve significantly. Notably, NatGen achieves a substantial increase of 0.1376 (0.7213-0.5837) in AUC and improves its F1 score from 0.5431 to 0.6735, indicating that it relies heavily on large-size data and struggles with generalization on smaller dataset.

\textbf{The trends of precision}: As the fine-tuning dataset size increases, most CodeLMs exhibit only a small improvement in precision, with a gentler growth trend compared to recall. Specifically, UniXcoder-base maintains a high precision across different dataset sizes, increasing from 0.6519 (20\% dataset) to 0.7020 (100\% dataset), a growth of 0.0501, making it the most stable CodeLM in terms of precision to dataset size. GraphCodeBERT-base and CodeBERT-base show relatively limited improvements in precision, rising from 0.6061 and 0.5822 to 0.6304 and 0.6008, with growths of only 0.0243 and 0.0186, respectively. CodeT5p-220M and NatGen have significantly lower precision on the 20\% dataset at just 0.5612 and 0.5250, respectively. Although their precision is improved on the 100\% dataset (reaching 0.6276 and 0.6043, respectively), their overall growths remain limited. This result suggests that increasing the dataset size has a limited impact on reducing false positives. Even with more training data, the growth of precision is constrained, likely due to the complexity of the CLB detection task, where the CodeLM still struggles to distinguish certain ambiguous cases.

\textbf{The trends of recall}: As the fine-tuning dataset size increases, all CodeLMs exhibit a noticeable improvement in recall, with most CodeLMs showing a much greater increase in recall compared to precision. For instance, UniXcoder-base's recall rises from 0.7857 (20\% dataset) to 0.8146 (100\% dataset), an increase of 0.0289. While the growth is modest, it consistently maintains a high level of performance across all dataset sizes. GraphCodeBERT-base and CodeBERT-base also show significant improvements, with recall increasing from 0.7143 and 0.7589 to 0.7754 and 0.7768, respectively, showing increases of 0.0611 and 0.0179. CodeT5p-220M and NatGen initially have lower recall at 20\% of the dataset (0.6964 and 0.5625), but as the dataset expands to 100\%, their recalls are improved to 0.7540 and 0.7607, respectively. Notably, NatGen sees the largest increase in recall, with a 0.1982 improvement, the highest among all CodeLMs. This trend indicates that increasing the dataset size significantly enhances the models' ability to identify positive samples, as seen by the substantial increase in recall. However, it is important to highlight that NatGen and CodeT5p-220M have much lower recall when the dataset is small, suggesting that these models are more dependent on larger datasets for higher performance.

In summary, the expansion of dataset size has a positive impact on the performance of fine-tuned CodeLMs in CLB detection. For the CLB detection tasks with limited data, UniXcoder-base demonstrates better data efficiency, achieving high performance even with small datasets. On the other hand, in scenarios with abundant data, CodeT5p-220M and NatGen benefit the most when fine-tuned for the CLB detection tasks, showing significantly enhanced performance as the dataset size increases.

\subsubsection{Impact of the Token Sequence Length}

\begin{figure}[ht]
    \centering
    \begin{minipage}[b]{0.49\textwidth}
        \centering
        \includegraphics[width=\linewidth]{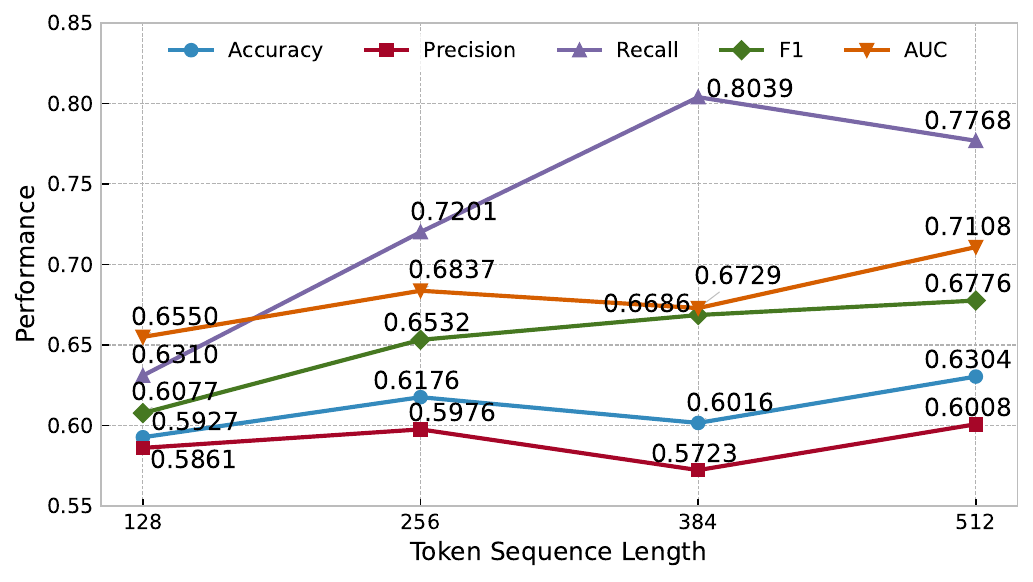}
        \subcaption{CodeBERT-base} \label{fig:sub1}
    \end{minipage}
    \hfill
    \begin{minipage}[b]{0.49\textwidth}
        \centering
        \includegraphics[width=\linewidth]{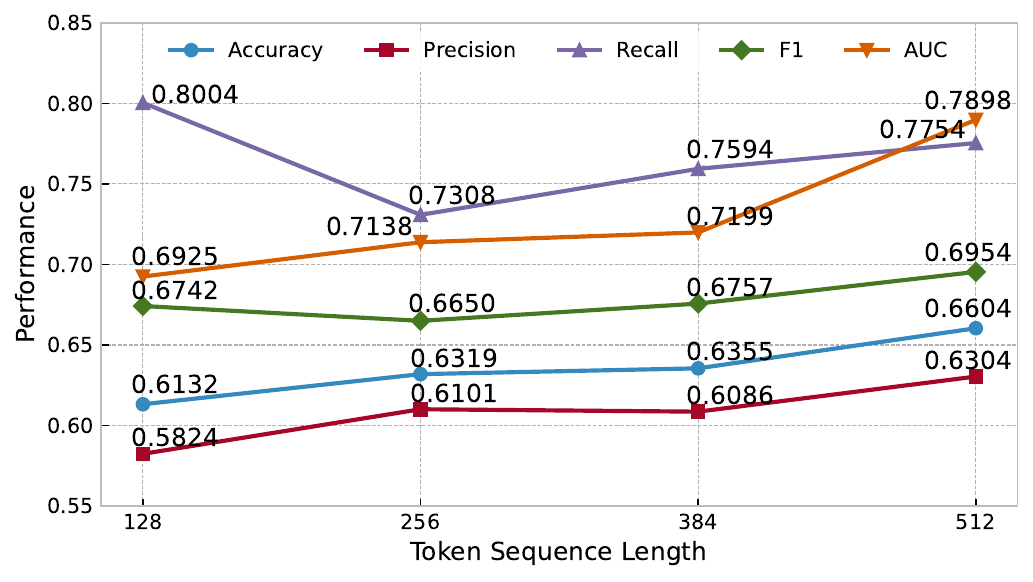}
        \subcaption{GraphCodeBERT-base} \label{fig:sub2}
    \end{minipage}

    \vspace{1em} 

    \begin{minipage}[b]{0.49\textwidth}
        \centering
        \includegraphics[width=\linewidth]{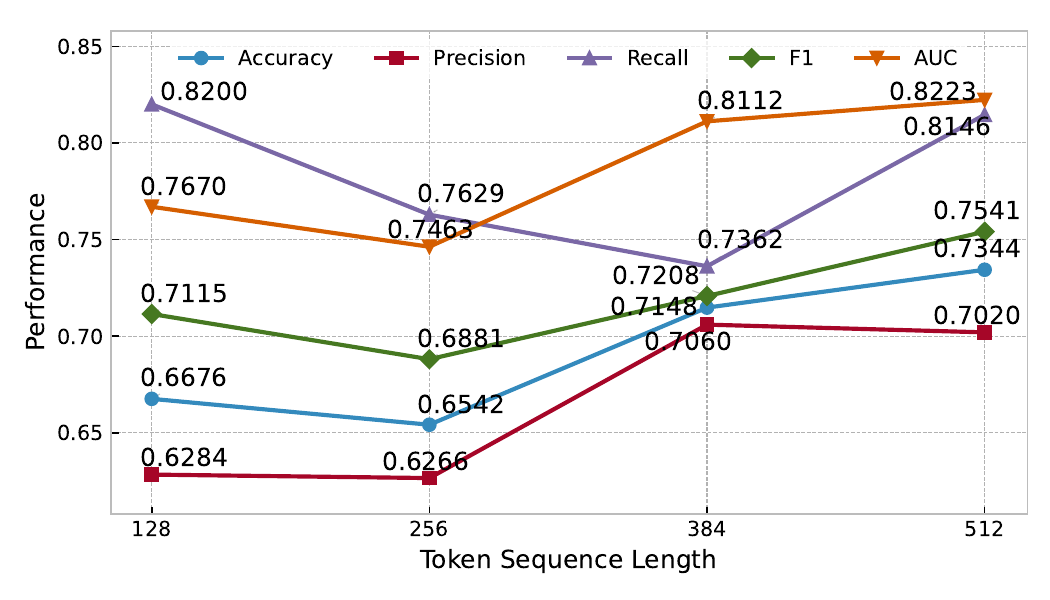}
        \subcaption{UniXcoder-base} \label{fig:sub3}
    \end{minipage}
    \hfill
    \begin{minipage}[b]{0.49\textwidth}
        \centering
        \includegraphics[width=\linewidth]{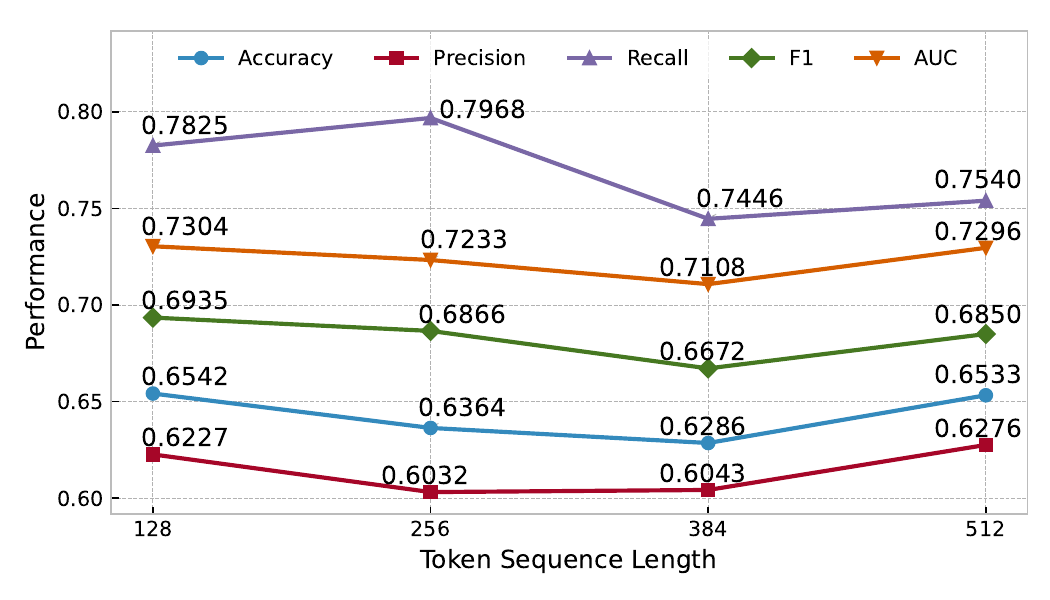}
        \subcaption{CodeT5p-220M} \label{fig:sub4}
    \end{minipage}

    \vspace{1em}

    \begin{minipage}[b]{0.49\textwidth}
        \centering
        \includegraphics[width=\linewidth]{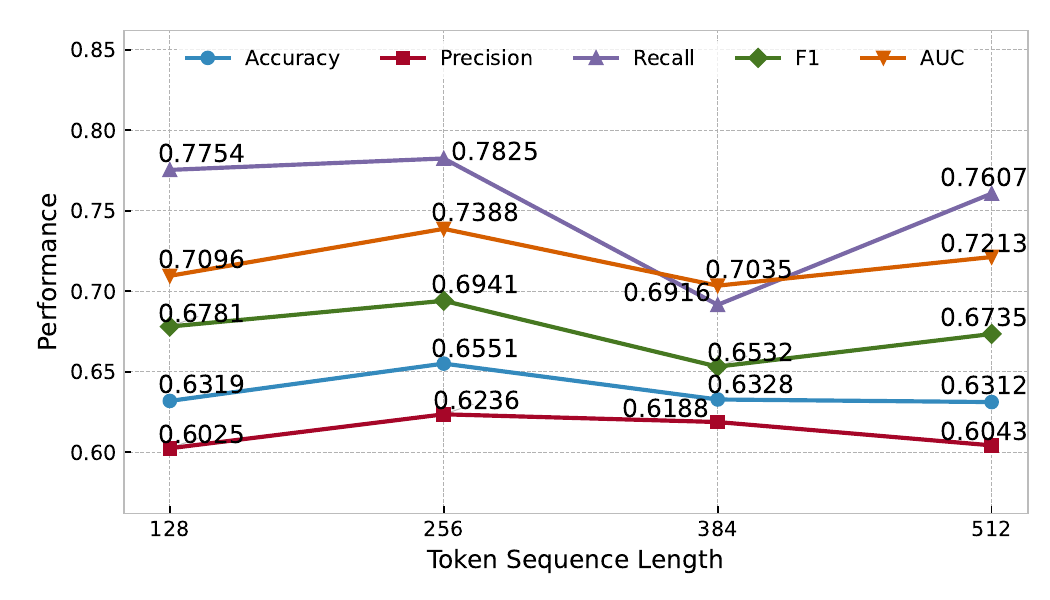}
        \subcaption{NatGen} \label{fig:sub5}
    \end{minipage}

    \caption{Performance of CodeLMs in detecting CLBs with different token sequence lengths}
    \label{fig:rq3_token}
\end{figure}

To investigate the impact of token sequence length on the performance of fine-tuned CodeLMs, we conducted experiments on five CodeLMs using various token sequence length (i.e., 128, 256, 384, and 512), and analyzed each CodeLM's performance across these configurations. The maximum token sequence length is 512 because this is the limit supported by some CodeLMs. For more details on CodeLMs, please refer to Table \ref{table:details_CodeLMs}. Figure \ref{fig:rq3_token} presents the detailed results of this experiment. Below is a specific analysis of the results:

\textbf{The overall trends of accuracy, F1, and AUC}: Overall, with the increase in token sequence length, the accuracy, F1, and AUC of most CodeLMs show an upward trend. UniXcoder-base leads in accuracy across all token sequence length settings, increasing from 0.6676 at the token sequence length of 128 to 0.7344 at 512, with F1 rising from 0.7115 to 0.7541, and AUC increasing from 0.7670 to 0.8223. This indicates that it can fully utilize longer token sequences to enhance overall performance in the CLB detection task. GraphCodeBERT-base also shows a noticeable growth trend, with its accuracy increasing from 0.6132 to 0.6604, F1 rising from 0.6742 to 0.6954, and AUC improving from 0.6925 to 0.7898, indicating its high sensitivity to input length constraints and its improved performance with extended token sequences. CodeBERT-base experiences a slight decrease in accuracy (0.6016) at the token sequence length of 384, but rebounds to 0.6304 at 512. F1 drops to 0.6532 at 256, then recovers to 0.6686 at 384 and 0.6776 at 512, indicating that the model has weaker sensitivity to short length token sequences but can gradually capture and leverage longer contexts. NatGen achieves the highest accuracy (0.6551) at 256, with a slight drop to 0.6328 at 384, and a further minimal change to 0.6312 at 512. The F1 trend follows a similar pattern to accuracy, which may indicate a learning bottleneck for the model at longer token sequence settings. The accuracy, F1, and AUC of CodeT5p-220M initially decrease and then increase as the token sequence length increases. However, the final results are not better than those at the token sequence length of 128, although the difference is small. This suggests that the model is not highly sensitive to changes in token sequence length.

\textbf{The trends of recall}: Recall of all CodeLMs generally increases with the increase in token sequence length, indicating that longer token sequences help the models capture more CLBs. GraphCodeBERT-base's recall decreases from 0.8004 at 128 to 0.7754 at 512, and CodeT5p-220M's recall reaches its highest value of 0.7968 at 256, then gradually decreases to 0.7540 at 512, suggesting that these two CodeLMs may encounter a decline in recall with very long token sequences. CodeBERT-base's recall steadily increases from 0.6310 at 128 to 0.8039 at 384, but drops to 0.7768 at 512, indicating that while recall can be improved with the increase in token sequence length, overly long inputs may cause interference in the CodeLM’s ability to understand code and detect CLBs due to redundant information. NatGen's recall reaches its highest value of 0.7825 at 256, but drops to 0.6916 at 384, then rebounds to 0.7607 at 512, showing some fluctuation.

\textbf{The trends of precision}: The trend of precision across different token sequence lengths is complex, with some CodeLMs showing an increase in precision as the token sequence length increases, while others show a decrease, suggesting that longer inputs may lead to some degree of misclassification. UniXcoder-base shows a steady improvement in precision across all token sequence length settings, increasing from 0.6284 at 128 to 0.7020 at 512, indicating that it can maintain high precision while improving recall. GraphCodeBERT-base starts with a precision of 0.5824 at 128, but it is improved to 0.6304 at 512. However, while its precision increases, its recall decreases, which may suggest that the CodeLM adopts a more conservative approach to reduce false positives, potentially leading to a lower recall. CodeT5p-220M and NatGen show relatively stable changes in precision. However, NatGen experiences a slight decline in precision at the token sequence length of 384 (0.6188) compared to the token sequence length of 256 (0.6236), while CodeT5p-220M's precision increases to 0.6276 at 512. CodeBERT-base shows the most noticeable drop in precision at 384 (0.5723), but it recovers to 0.6008 at 512, further indicating its weaker adaptability to longer sequences.

In summary, UniXcoder-base performs best across all token sequence length settings, making it suitable for tasks with high overall performance requirements. The token sequence length has a significant impact on the performance of fine-tuned CodeLMs in CLB detection. Different CodeLMs exhibit varying responses to changes in token sequence length. Some CodeLMs perform better with shorter token sequence lengths, while others show improved performance with longer token sequence lengths.

\begin{tcolorbox}
\textbf{Finding 3}: Increasing the size of the fine-tuning dataset for CodeLMs can indeed improve the model's performance in CLB detection to some extent. However, within the scope of our experimental setup, increasing the token sequence length does not necessarily lead to better performance. Different CodeLMs show varying sensitivity to token sequence length; some CodeLMs (e.g., CodeT5p-220M) perform exceptionally well with shorter token sequence lengths, and their performance decreases as the token sequence length increases. On the other hand, some CodeLMs (e.g., GraphCodeBERT-base) perform better with longer token sequence lengths. 
\end{tcolorbox}

\subsection{RQ4: Impact of Code Comments on Fine-tuned CodeLM Performance}\label{RQ4Results}

We have already conducted fine-tuning experiments and analysis on the CLB dataset without code comments, and the results are shown in Table \ref{table:rq1}. To answer RQ4, we carried out fine-tuning experiments on the CLB dataset with code comments, and the results are shown in Table \ref{table:rq4}. In addition, we include GPT-4o-mini as an inference-only reference baseline evaluated on the same commented test set using prompt-based reasoning, without any task-specific fine-tuning or parameter updates.
By comparing the performance of CodeLMs fine-tuned on the CLB dataset with and without comments, we analyze the impact of code comments on the performance of CodeLMs in the following two aspects. 

\begin{table}[ht]
\centering
\renewcommand\arraystretch{1.2}
\caption{Performance of CodeLMs in detecting CLBs after fine-tuning on the CLB dataset with code comments}
{
\begin{tabular}{lccccc}
\hline
\textbf{CodeLM} & \textbf{Acc} & \textbf{Prec} & \textbf{Rec} & \textbf{F1} & \textbf{AUC} \\ \hline
GPT-4o-mini (Baseline) & \grayA{0.5152} & \grayA{0.5171} & \grayA{0.4589} & \grayA{0.4863} & \grayA{0.5152} \\
CodeBERT-base & \grayA{0.6304} & \grayA{0.6008} & \grayA{0.7768} & \grayA{0.6776} & \grayA{0.7108} \\
GraphCodeBERT-base & \grayA{0.6604} & \grayA{0.6304} & \grayA{0.7754} & \grayA{0.6954} & \grayA{0.7898} \\
UniXcoder-base & \grayA{0.7344} & \grayA{0.7020} & \grayA{0.8146} & \grayA{0.7541} & \grayA{0.8223} \\
CodeT5-base & \grayA{0.6096} & \grayA{0.5714} & \grayA{0.8770} & \grayA{0.6920} & \grayA{0.7046} \\
CodeT5p-220M & \grayA{0.6533} & \grayA{0.6276} & \grayA{0.7540} & \grayA{0.6850} & \grayA{0.7296} \\
CodeT5p-770M & \grayA{0.5407} & \grayA{0.5102} & \grayA{0.7756} & \grayA{0.6159} & \grayA{0.6043} \\
NatGen & \grayA{0.6312} & \grayA{0.6043} & \grayA{0.7607} & \grayA{0.6735} & \grayA{0.7213} \\
Deepseek-Coder-1.3B-base & \grayA{0.5482} & \grayA{0.5250} & \grayA{0.7673} & \grayA{0.6235} & \grayA{0.5965} \\
Granite-3B-Code-base-2k & \grayA{0.5075} & \grayA{0.4894} & \grayA{0.5584} & \grayA{0.5216} & \grayA{0.5587} \\
StarCoder2-3B & \grayA{0.5696} & \grayA{0.5443} & \grayA{0.7197} & \grayA{0.6198} & \grayA{0.6049} \\
CodeQWen1.5-7B & \grayA{0.5280} & \grayA{0.5048} & \grayA{0.9591} & \grayA{0.6615} & \grayA{0.5471} \\
Qwen2.5\_Coder-7B & \grayA{0.5468} & \grayA{0.5300} & \grayA{0.5064} & \grayA{0.5180} & \grayA{0.5915} \\
CodeLlama-7b-hf & \grayA{0.5160} & \grayA{0.5242} & \grayA{0.5675} & \grayA{0.5450} & \grayA{0.5402} \\ \hline
\end{tabular}
}

\label{table:rq4}
\end{table}

\subsubsection{The general impact of code comments on fine-tuned CodeLMs performance}

Compared to the fine-tuning results without code comments, most CodeLMs showed an overall performance improvement after fine-tuning with commented data, indicating that code comments provide additional contextual information, helping the model better identify potential CLBs. For UniXcoder-base, all metrics were improved after the inclusion of code comments, suggesting that the addition of code comments significantly enhanced the model's performance in CLB detection, especially in terms of accuracy, F1 score, and AUC.
GraphCodeBERT-base, after fine-tuning with code comments, saw an improvement in AUC from 0.7447 to 0.7898 and in precision from 0.6128 to 0.6304. Although recall dropped slightly (from 0.8241 to 0.7754), the increases in AUC and precision indicate that the inclusion of code comments helped improve the model's performance of CLB detection.
CodeT5-base, on the other hand, saw a substantial increase in recall from 0.6840 to 0.8770 and a moderate improvement in F1 score from 0.6586 to 0.6920. While accuracy and precision decreased, this indicates that code comments helped the model enhance its overall CLB detection capabilities. However, this improvement came at the cost of a slight reduction in precision.
For CodeT5p-220M, fine-tuning with code comments resulted in a noticeable boost across several metrics: accuracy improved from 0.6364 to 0.6533, precision from 0.6148 to 0.6276, recall from 0.7307 to 0.7540, and F1 score from 0.6678 to 0.6850. This result shows that code comments contributed to an overall performance improvement for this model.
For reference, GPT-4o-mini achieves moderate performance on the commented CLB dataset under inference-only settings. However, its overall performance remains consistently lower than that of fine-tuned CodeLMs, indicating that the availability of code comments alone is insufficient to fully compensate for the absence of task-specific adaptation.

Although most CodeLMs show improvement after fine-tuning with code comments, some CodeLMs fail to benefit from the code comments and even experience performance degradation. For instance, CodeT5-base and Granite-3B-Code-base-2k exhibit a noticeable decline in accuracy and precision. For Granite-3B-Code-base-2k, the recall also significantly decreased. The precision drop, leading to a decrease in AUC and F1 score, results in that the overall performance does not meet expectations.
After fine-tuning CodeQwen1.5-7B with code comments, the AUC decreased from 0.5706 to 0.5471. Despite a significant increase in recall (from 0.7580 to 0.9591), the drop in AUC suggests that while code comments helped the CodeLM identify more CLB instances, they also weakened the model’s discriminative ability, resulting in an overall decline in performance.

\subsubsection{Impact of code comments on recall and precision}

Code comments significantly improve recall, which is crucial for CLB detection tasks, as increasing recall helps reduce false negatives and capture more potential bugs. However, in some CodeLMs, this often comes with a decrease in precision. For example:
The precision of CodeT5-base notably dropped from 0.6350 to 0.5714, despite a substantial increase in recall (from 0.6840 to 0.8770). This phenomenon suggests that the inclusion of code comments might lead to more false positives, as the boost in recall is accompanied by a significant decline in precision.
Compared with the fine-tuning results without code comments, Granite-3B-Code-base-2k's accuracy dropped from 0.5516 to 0.5075, and precision decreased from 0.5275 to 0.4894. Despite an increase in recall from 0.6951 to 0.7197, the decline in precision and F1 score (from 0.5998 to 0.5584) indicate that the inclusion of code comments did not result in an overall performance improvement.
For Deepseek-Coder-1.3B-base, after fine-tuning with code comments, the recall increased from 0.7063 to 0.7673, and the F1 score rose from 0.6026 to 0.6235. Although both the recall and F1 score were improved, the AUC dropped from 0.6125 to 0.5965, with the accuracy and precision showing little change. This suggests that code comments may have led to more misclassifications, reducing the CodeLM's discriminative ability.
A similar tendency can be observed from the inference-only GPT-4o-mini baseline, whose recall does not substantially benefit from the presence of code comments. This observation further suggests that effectively leveraging code comments for CLB detection requires model adaptation through fine-tuning, rather than relying solely on prompt-based reasoning.

\begin{tcolorbox}

\textbf{Finding 4}: Incorporating code comments into the CLB dataset generally improves CodeLM performance, particularly in recall and F1 score, enabling the detection of more potential CLBs. However, such improvements may come at the cost of reduced precision for some models, leading to limited or even negative overall performance gains. Moreover, results from the inference-only GPT-4o-mini baseline suggest that code comments alone are insufficient to guarantee improved CLB detection performance, underscoring the necessity of task-specific fine-tuning to effectively leverage comment-level information.

\end{tcolorbox}

\section{Discussion}\label{chap:discussion}

\subsection{Interpretation of Study Results}

\textbf{RQ1: CodeLM performance after fine-tuning}: This study reveals significant performance differences among CodeLMs of various sizes in CLB detection tasks. Overall, the evaluated CodeLMs exhibited varying levels of performance after task-specific adaptation, with some models achieving competitive results while others remained relatively modest. This observation suggests that the CLB dataset we constructed effectively enhances CodeLMs' modeling capabilities in handling CLB detection challenges.

Notably, the size of model parameters exhibits a considerable impact on the performance of CodeLMs fine-tuned for CLB detection. As shown in Tables \ref{table:details_CodeLMs} and \ref{table:rq1}, under our experimental setup, CodeLMs with a relatively smaller number of parameters (no more than 220M) generally achieved better results, while large CodeLMs, such as Qwen2.5\_Coder-7B and CodeLlama-7b-hf, demonstrated only moderate performance. It is important to note that this observation is made under specific conditions, where smaller models are fully fine-tuned, several larger models are adapted using LoRA, and all models are evaluated with a maximum input length of 512 tokens. Therefore, within this setting, solely increasing the model size does not necessarily guarantee superior performance in CLB detection tasks. We speculate that under these constraints, small CodeLMs, due to their more streamlined architecture and more focused training, may be more amenable to fine-tuning and adaptation for targeted tasks such as CLB detection~\cite{xu2025evaluating, hsieh2023distilling,ahouzi2024cfm}.

In addition, the inclusion of GPT-4o-mini as an inference-only reference baseline provides complementary insights into the role of task-specific adaptation. As an inference-only general LLM, GPT-4o-mini is consistently outperformed by fine-tuned CodeLMs across all evaluation metrics. This result indicates that, under our experimental setup, prompt-based reasoning alone, without parameter adaptation, may be insufficient to fully capture the complex cross-language interaction patterns required for accurate CLB detection.

Furthermore, UniXcoder-base is ranked first across all evaluation metrics, highlighting that the depth of a CodeLM’s understanding of code structure and semantics during pre-training plays a crucial role in cross-language tasks. 
The contrast between fine-tuned CodeLMs and the inference-only GPT-4o-mini baseline further suggests that effective CLB detection in this setting may benefit more from task-specialized representation learning than from prompt-based general-purpose reasoning capabilities alone. However, we interpret this comparison cautiously. As a relatively compact and earlier-generation model, GPT-4o-mini serves as a helpful reference point for lightweight inference but does not represent the capabilities of state-of-the-art frontier LLMs (e.g., GPT-4 or Claude 3.5 Sonnet). Consequently, this baseline illustrates the performance gap between non-adapted general models and fine-tuned domain models of similar scales, but it does not reflect the detection capabilities of current state-of-the-art models on the CLB dataset.
These findings suggest that, in our setting, the pre-training strategy and architectural design of a CodeLM may play a more important role than merely scaling up the number of parameters, offering valuable insights for the future design of efficient CodeLMs.

\textbf{RQ2: Impact of Different Fine-Tuning Datasets on CLB Detection Performance}: The results of this RQ suggest that fine-tuning CodeLMs on single-language bug datasets provides limited benefits for CLB detection.
While such datasets are effective for learning language specific bug patterns, they do not adequately support capturing the characteristics of cross-language interactions required for CLB detection.

This limitation can be attributed to the fundamental differences between single-language bugs and CLBs. Single-language bug datasets mainly capture defects arising from syntax, semantics, or control flow within an individual PL. In contrast, CLBs typically originate from mismatches and incorrect assumptions at language boundaries, such as inconsistent data representations, interface misuse, or incompatible interaction protocols between multiple PLs. These cross-language characteristics are inherently absent from single-language training data, making it difficult for CodeLMs to learn transferable representations for CLB detection.

The findings highlight that effective CLB detection depends not only on CodeLM capacity but also on the relevance of the training data. Fine-tuning on CLB dataset allows CodeLMs to directly observe and learn interaction patterns across PLs, which is essential for detecting CLBs. Therefore, relying solely on single-language bug datasets is insufficient for addressing the unique challenges posed by cross-language software systems.

To further examine whether CodeLMs fine-tuned on the CLB dataset can transfer to single-language bug detection tasks, we conduct experiments on two single-language datasets, CVEfixes and CodeNet, together with our CLB dataset. Specifically, for each single-language dataset, we fine-tune two CodeLMs on the CLB dataset and on the target single-language dataset, respectively. We then evaluate all fine-tuned CodeLMs on the test set of the corresponding single-language datasets. This experimental setting is consistent with that used in RQ2.

\begin{table*}[ht]
\centering
\renewcommand\arraystretch{1.2}
\caption{Single-language Bugs Detection Performance of CodeLMs Fine-Tuned on CVEfixes and the CLB Dataset}
\resizebox{\textwidth}{!}{
{
\begin{tabular}{lcccccccccc}

\hline
\multicolumn{1}{c}{\multirow{2}{*}{\textbf{CodeLM}}} & \multicolumn{5}{c}{\textbf{Fine-tuned in CLB}} & \multicolumn{5}{c}{\textbf{Fine-tuned in CVEfixes}} \\ \cline{2-11} 
\multicolumn{1}{c}{} & \textbf{Acc}    & \textbf{Prec}   & \textbf{Rec}    & \textbf{F1}     & \textbf{AUC}    & \textbf{Acc}    & \textbf{Prec}   & \textbf{Rec}    & \textbf{F1}     & \textbf{AUC}    \\ \hline
CodeBERT-base   & \grayA{0.4923} & \grayA{0.4908} & \grayA{0.4065} & \grayA{0.4447} & \grayA{0.4776} & \grayA{0.6352} & \grayA{0.6000} & \grayA{0.8112} & \grayA{0.6898} & \grayA{0.6867} \\
UniXcoder-base  & \grayA{0.5051} & \grayA{0.5027} & \grayA{0.9388} & \grayA{0.6548} & \grayA{0.4829} & \grayA{0.6480} & \grayA{0.6033} & \grayA{0.8639} & \grayA{0.7105} & \grayA{0.7021} \\
\hline
\end{tabular}
}
}
\label{table:cvefixes_dis}
\end{table*}

\begin{table*}[ht]
\centering
\renewcommand\arraystretch{1.2}
\caption{Single-language Bugs Detection Performance of CodeLMs Fine-Tuned on CodeNet and the CLB Dataset}
\resizebox{\textwidth}{!}{
{
\begin{tabular}{lcccccccccc}

\hline
\multicolumn{1}{c}{\multirow{2}{*}{\textbf{CodeLM}}} & \multicolumn{5}{c}{\textbf{Fine-tuned in CLB}} & \multicolumn{5}{c}{\textbf{Fine-tuned in CodeNet}} \\ \cline{2-11} 
\multicolumn{1}{c}{} & \textbf{Acc}    & \textbf{Prec}   & \textbf{Rec}    & \textbf{F1}     & \textbf{AUC}    & \textbf{Acc}    & \textbf{Prec}   & \textbf{Rec}    & \textbf{F1}     & \textbf{AUC}    \\ \hline
CodeBERT-base   & \grayA{0.4965} & \grayA{0.4982} & \grayA{0.9890} & \grayA{0.6626} & \grayA{0.4421} & \grayA{0.6605} & \grayA{0.6341} & \grayA{0.7591} & \grayA{0.6909} & \grayA{0.7214} \\
UniXcoder-base  & \grayA{0.5070} & \grayA{0.5036} & \grayA{0.9820} & \grayA{0.6658} & \grayA{0.4992} & \grayA{0.6745} & \grayA{0.6480} & \grayA{0.7640} & \grayA{0.7012} & \grayA{0.7290} \\
\hline

\end{tabular}
}
}
\label{table:codenet_dis}
\end{table*}

The experimental results are shown in Tables \ref{table:cvefixes_dis} and \ref{table:codenet_dis}. When evaluated on single-language benchmarks, models fine-tuned on the CLB dataset achieve accuracy and AUC values close to random guessing. In contrast, models fine-tuned on single-language datasets consistently obtain higher accuracy, F1 score, and AUC on both CodeBERT-base and UniXcoder-base. These results indicate that models fine-tuned on the CLB dataset have limited discriminative capability for single-language bug detection. This performance gap further suggests that features learned from cross-language interactions do not transfer well to bug detection within a single programming language.

Moreover, the performance degradation on single-language benchmarks indicates that CLB fine-tuning does not improve model effectiveness beyond its intended task. Instead, fine-tuning on the CLB dataset mainly guides CodeLMs to learn CLB features, which are largely unrelated to the characteristics of single-language bugs. Overall, this study emphasizes the necessity of constructing dedicated CLB datasets, which enable CodeLMs to learn CLB characteristics and interaction patterns essential for effective CLB detection.

\textbf{RQ3: Performance influencing factors of fine-tuned CodeLMs}: Through our experiments, we observed that the size of the dataset used during the fine-tuning stage has a positive impact on CodeLMs' performance. As the dataset size gradually increases, all five evaluated CodeLMs exhibited varying degrees of improvement in CLB detection tasks. However, different CodeLMs showed differing sensitivities to changes in dataset size. Some CodeLMs experienced rapid performance gains with slight increases in dataset size, while others demonstrated more stable or gradual trends. Further analysis reveals that this variation primarily stems from the following two factors:
First, the alignment between the pre-training objectives and the CLB detection task greatly influences data efficiency. For instance, UniXcoder, which adopts multimodal pre-training objectives (e.g., code-text alignment and AST path modeling), can generalize well even under low-resource conditions. UniXcoder-base effectively captures CLB patterns, enabling fast convergence and strong performance. In contrast, generative models like NatGen, which rely heavily on code generation and reconstruction tasks during pre-training, require a larger number of samples to learn input-output mappings. As a result, both UniXcoder-base and NatGen exhibit weaker generalization in low-resource scenarios.
Second, model parameter size also affects data utilization efficiency. Small models tend to reach their performance upper bound with limited data, while large models require substantial amounts of data to fully activate their potential. However, this also comes with higher training and inference costs. Therefore, the sensitivity of CodeLMs to dataset size during fine-tuning is not only influenced by their architectural design and parameter configuration but also reflects the effectiveness of semantic transfer between pre-training and downstream tasks.

Regarding the impact of token sequence length, we found that increasing the token sequence length could improve performance to a certain degree, but the performance gains were not consistently monotonic. Some CodeLMs, such as UniXcoder-base and GraphCodeBERT, employ relative position encoding, which allows better generalization to longer sequences. These CodeLMs show stable performance gains as the token sequence length increases, particularly at the length of 512, where they can more effectively leverage contextual information to improve structural understanding.
In contrast, CodeLMs like CodeBERT-base and NatGen, which use absolute position encoding, experience performance degradation when the input token sequence length exceeds the pre-training maximum. This suggests limitations in their ability to generalize positional information. Moreover, architectural characteristics also influence how models respond to longer token sequences. For example, GraphCodeBERT-base can recover more complete data flow structures with longer inputs, and UniXcoder-base strengthens structural reasoning by linearizing AST paths in extended contexts.

These findings suggest that, in practical applications, the choice of token sequence length should be determined by both the characteristics of the dataset and the model’s ability to process long token sequences, rather than pursuing longer token sequence length blindly.

\textbf{RQ4: Impact of code comments on fine-tuned CodeLM performance}: In most cases, including code comments significantly improved the CodeLMs' recall and F1 scores on the CLB detection task, suggesting that comments help CodeLMs better understand code semantics and capture potential bugs. However, for some CodeLMs, the inclusion of code comments led to a decrease in precision, and the gains in recall were insufficient to offset this decrease in precision, resulting in an overall drop in performance.
Further analysis revealed that this phenomenon is primarily related to the constraint of fixed token sequence length. After incorporating code comments, the amount of code information within a fixed-length token sequence was reduced, preventing the model from accessing complete information and thus lowering its prediction accuracy.

For CodeLMs that showed notable performance improvements, code comments provided additional context and explanations that contributed to a more accurate understanding of the code, suggesting that these CodeLMs possess relatively limited inherent code understanding capabilities.
In contrast, the inference-only GPT-4o-mini baseline did not consistently benefit from the inclusion of code comments, suggesting that, for this non-fine-tuned reference setting, comment-level semantic information alone 
may be insufficient to substantially improve CLB detection performance without task-specific fine-tuning.
Therefore, in practical applications, whether to retain code comments should be flexibly decided based on the specific task requirements. It is also advisable to consider adopting more flexible token budget strategies, such as moderately increasing the token sequence length limit or prioritizing key code content. The results from GPT-4o-mini additionally suggest that, even for relatively strong general-purpose models, input design alone may not fully compensate for the lack of task-aligned training. Nevertheless, this comparison should be interpreted cautiously, as GPT-4o-mini is used here only as a prompt-based reference baseline rather than as a representative upper bound of current frontier model performance. This finding further indicates that current CodeLMs still have considerable potential for improvement in code understanding tasks involving cross-language interactions.

The results indicate that, in future input design for CodeLMs, a more systematic optimization of the weighting between code information and auxiliary semantic information, such as comments, is necessary to maximize the overall model performance.

\subsection{Implications}
Based on the study results and their analysis, we obtained a number of implications for practitioners and researchers, respectively. All implications below are derived under the experimental settings of this study, including the selected open-source CodeLMs, the constructed CLB dataset, and the fine-tuning and evaluation protocols described in Section~\ref{chap:expdesign}.

\subsubsection{Implications for practitioners}
\textbf{(1) Emphasizing the specific characteristics of CLBs.}
This study shows that CLBs are not only considerable in quantity but also tend to be more obscure and harder to detect compared to single-language bugs, due to the involvement of interaction mechanisms across different languages.
This observation is supported by the consistently lower detection performance obtained when CodeLMs are fine-tuned only on single-language bug datasets (RQ2), indicating that CLBs exhibit characteristics that are insufficiently captured by single-language bug patterns alone.
Therefore, for projects that involve cross-language interactions similar to those represented in our CLB dataset, it is advisable to explicitly consider risks related to cross-language interaction during development and testing, such as data type mismatches, inconsistent exception handling, and cross-language resource management.

\textbf{(2) Paying attention to limitations of CodeLMs in CLB detection.}
Although CodeLMs have made significant progress in tasks such as code understanding and generation, their performance on CLB detection remains limited.
As evidenced by the experimental results in RQ1 and RQ2, the detection performance of the evaluated fine-tuned CodeLMs is highly dependent on the availability of CLB-specific training data and does not consistently generalize from single-language bug datasets.
Practitioners should not regard the outputs of CodeLMs as final conclusions when detecting CLBs. Instead, a multifaceted and multi-perspective detection process should be adopted by combining static analysis tools, rule-based detection methods, and manual review to comprehensively assess the possibility of CLBs. Over-reliance on the output of a single CodeLM should be avoided to enhance overall reliability of CLB detection.

\textbf{(3) Considering the impact of code comments under fixed token length constraints.}
In this study, the results show that including code comments generally improves recall and F1 score for most CodeLMs.
However, all experiments were conducted under a fixed maximum token sequence length (up to 512 tokens).

Our results therefore only indicate that, under fixed-length input constraints, the inclusion of comments provides additional contextual information but may also lead to information loss when input context is truncated.
In such cases, the model is forced to make predictions based on incomplete input, which can negatively affect detection accuracy.
Future work will explore the performance of CodeLMs that support longer context lengths under more powerful computational resources, in order to better capture complex cross-language interactions without aggressive truncation.

\textbf{(4) Selecting an appropriate CodeLM based on resource availability and projects under detection.}
Under the specific experimental conditions of this study, several small CodeLMs (e.g., UniXcoder-base and GraphCodeBERT-base) achieved better or comparable performance to the evaluated larger models. It suggests that the potential advantages of large models may not be fully realized under constrained input lengths and limited fine-tuning data, which are particularly challenging in cross-language scenarios involving additional boilerplate and interaction code.
Accordingly, within similar resource and input constraints, smaller CodeLMs may represent a cost-effective choice, while the effectiveness of larger CodeLMs with extended context lengths remains an open question for future investigation.

\subsubsection{Implications for researchers}

\textbf{(1) Expanding the coverage of CodeLMs and tasks.}
The current study evaluates a limited number of CodeLMs, with different model architectures and parameter sizes. Future work should systematically expand the research scope to include a broader range of CodeLM architectures and large CodeLMs. Additionally, the transferability and robustness of CodeLMs across diverse cross-language scenarios should be explored to establish more comprehensive and accurate evaluation standards. Future work should also evaluate the CLB detection performance of frontier models on our dataset under zero-shot or few-shot settings.

\textbf{(2) Building larger and more diverse CLB dataset.}
This study provides evidence of the importance of dataset quality and fine-tuning strategies. However, existing CLB datasets are still limited in terms of size, language diversity, and interaction patterns. Future work should focus on constructing larger and more comprehensive benchmark datasets that cover a wider range of PL combinations and cross-language interaction modes, in order to systematically evaluate the detection capabilities of CodeLMs of different scales and structures.

\textbf{(3) Investigating the fundamental characteristics and classification of CLBs.}
Currently, the understanding of CLBs remains relatively shallow, and there is a lack of systematic classification standards regarding the various types of CLBs. Future research could aim to establish a structured classification system for CLBs based on their sources (e.g., memory management, exception handling, data format conversion) and manifestations (e.g., data loss, type errors, resource leaks). Such a classification system would not only facilitate more targeted CodeLM training but also support the development of finer-grained CLB detection strategies.

\textbf{(4) Exploring fine-tuning strategies and data augmentation methods for CLB detection.}
While it is shown that increasing the volume of training data has improved CodeLM performance, efficiently leveraging limited data remains a significant challenge. Future directions include exploring self-supervised learning, domain adaptation, and cross-language contrastive learning to enhance model performance under low-resource conditions. Additionally, designing smarter data augmentation methods, such as automatically generating diverse CLB cases, could further enrich the training dataset and ultimately improve the CodeLMs’ performance in detecting CLBs.

\textbf{(5) Exploring the integration of static analysis and deep learning models for CLB detection.}
While this study focuses on data-driven models, traditional static analysis remains a crucial approach that struggles with language boundaries. Inspired by recent neural and symbolic frameworks such as IRIS \cite{li2025iris}, future research may explore integrating deep learning with static analysis for CLB detection. By combining the rigor of traditional static analysis with the semantic understanding of deep learning models across languages, this hybrid approach presents a highly promising direction for developing advanced CLB and vulnerability detection systems.

\section{Threats to Validity}\label{chap:threats}

\subsection{Construct Validity}
A primary threat to construct validity arises from the fact that our dataset relies on CLCFinder, a tool designed to detect cross-language code, and the data collection tool we developed, as there is currently no specialized tool to identify cross-language code. This may introduce inaccuracies in the identification of cross-language code, thereby affecting the effectiveness of the instances in the dataset. To mitigate this threat, we recognized cross-language code only through nine cross-language interaction mechanisms that span three language combinations, ensuring that the identified code engages in valid cross-language interaction. This design helps improve the accuracy of cross-language code identification and enhances the reliability of dataset construction.

Another potential threat lies in the data collection process, particularly in the selection of open-source projects from GitHub, which may introduce selection bias. Since our dataset is derived from popular and actively maintained repositories, this biased sampling may impact the representativeness of the collected data and thus affect the generalizability of our findings. To reduce this threat, we provided a detailed description of our data collection procedure in Section~\ref{data_collection}, drawing inspiration from the approach of Li et al.~\cite{li2022polyfax} to enhance methodological transparency and facilitate comparison with existing studies in the field. 

\subsection{External Validity}
To answer RQ1 and RQ4, we fine-tuned CodeLMs on datasets with and without code comments, respectively, to investigate their performance in CLB detection tasks and to analyze the impact of code comments on CodeLMs effectiveness. A major threat to external validity in this setting lies in the limited parameter sizes of the CodeLMs used. In this study, we fine-tuned only 13 relatively small CodeLMs (no larger than 7B). As a result, the findings may not generalize to the performance of large models in CLB detection. Furthermore, due to computational constraints, we employed LoRA, a parameter-efficient fine-tuning (PEFT) technique, for CodeLM variants with more than 770M parameters. This may affect the generalizability of our study findings. In addition, GPT-4o-mini was included in RQ1 and RQ4 merely as an inference-only reference baseline for prompt-based, non-fine-tuned performance. Therefore, its results should not be interpreted as representing the upper bound of current frontier LLM performance, and the corresponding comparisons should be generalized with caution.

To answer RQ2, we fine-tuned CodeLMs on both single-language bug and CLB datasets to explore whether single-language bug data can support CLB detection. The validity of this investigation is also threatened by the limited coverage of datasets. In this task, we limited the datasets to two bug detection datasets that contain both Java and Python for the sake of experimental consistency. We did not include more extensively studied single-language bug datasets, such as those for C/C++. This constraint may limit the applicability of our findings to broader real-world settings. Nevertheless, we selected datasets with consistent language to ensure comparability and experimental consistency under our current research conditions.

To answer RQ3, we designed experiments to analyze the effects of fine-tuning dataset size and input token sequence length on CodeLMs performance. In this task, we focused on the impact of token sequence length below 512, and thus we could not yet assess the effect of longer token sequence on CodeLMs performance in CLB detection tasks. Regarding the impact of data volume, our analysis was limited to datasets of less than 11,126 samples (5,563 pairs), which is relatively small for LLMs. Whether larger datasets would enhance the performance of CodeLMs warrants further investigation.

\subsection{Internal Validity}
A key internal threat lies in the manual analysis of CLB symptoms and root causes may introduce internal validity threats due to reliance on human judgment. To mitigate this issue, we followed an established taxonomy from prior work with minor adaptations to better fit the characteristics of the CLB dataset. Moreover, all sampled cases were independently analyzed by the first and second authors, and disagreements were resolved through discussion to reach consensus, thereby reducing subjectivity and improving internal validity.

Another important internal threat concerns the choice of hyper-parameters during the fine-tuning process. The hyper-parameters of these 13 CodeLMs during the fine-tuning process play a crucial role in the final performance on the CLB detection task. Changes in hyper-parameters can lead to different results, potentially affecting the reliability of our experimental findings. To mitigate this threat, we conducted multiple test experiments to determine the optimal hyper-parameters.

A further internal threat stems from the different fine-tuning strategies used for CodeLMs of different sizes: small CodeLMs were fine-tuned using full training, while large CodeLMs were fine-tuned using PEFT. This inconsistency may affect the fairness of model comparisons. To alleviate this threat, we explicitly reported the fine-tuning strategy used for each model in our results, and we highlighted these differences in our comparative analysis. When interpreting performance differences, we took the fine-tuning method into account to avoid drawing misleading conclusions based on absolute scores.

Additionally, data leakage is one of the internal validity threats in our study, which is a common issue in CodeLM research. The CLB dataset we collected was sourced from publicly available GitHub repositories and there is a potential overlap with the pre-training data of some CodeLMs used in our study. Some data in the CLB dataset may predate the release of certain CodeLMs, and the training datasets for some of these models are not explicitly disclosed. This means that we cannot fully eliminate potential data leakage simply by filtering based on CodeLM release dates or the CodeLMs' training dataset information. However, as noted by Guo et al. \cite{guo2024exploring}, the presence of overlapping or contaminated data does not necessarily affect model evaluation, and models may even perform better in some cases on data that postdates their training cutoff date. 


\subsection{Reliability}
A key threat to reliability stems from potential issues in the internal consistency and stability of the software tools used for cross-language code identification and data collection. Since these tools play a central role in dataset construction, any errors or data processing biases in their implementation could lead to unreliable datasets. To mitigate this threat, the second author took primary responsibility for developing the tools, while the first author conducted regular reviews of the key modules. Extensive unit and integration tests were performed to ensure functional correctness and robustness. This process helps to ensure that consistent results are obtained under similar conditions, thereby enhancing the reliability of our findings.

To further ensure the reliability and reproducibility of our experiments, we publicly shared the data collection standards and the final CLB dataset \cite{dataset}. Detailed descriptions were provided for each step of the data collection process (see Section \ref{data_collection}), including repository selection criteria (e.g., repositories with more than 500 stars, cross-language collaboration, and relevant bug issues) and the procedure of data processing. These steps ensured the standardization of data collection, and all relevant parameters (such as the repository star threshold and the minimum proportion of code in each PL) were clearly specified and justified in the methodology. 

In the dataset partitioning phase, we used a fixed random seed to ensure consistent splitting of the training, validation, and test sets, thereby eliminating randomness that could affect the results. Additionally, we made our experimental scripts publicly available\cite{dataset} to guarantee the reproducibility of our experiments. Thus, with these measures, the threats to reliability are alleviated.

\section{Conclusions and Future Work}\label{chap:conclusions}
In this work, we designed CLCFinder, a tool for identifying cross-language code, and constructed a real-world dataset for CLB detection from GitHub. Based on this dataset, we conducted experiments to evaluate the performance of widely-used CodeLMs in the CLB detection task. The results show significant performance variation across different CodeLMs, with small CodeLMs (the parameter size is 220M or less) tending to achieve better performance than large CodeLMs within our experimental setting. To further validate the differences between our CLB dataset and single-language bug datasets, we fine-tuned CodeLMs on SOTA single-language bug detection datasets and applied these fine-tuned CodeLMs to CLB detection. The results demonstrate significant differences in feature characteristics between single-language bug and CLB datasets, validating the effectiveness and necessity of our CLB dataset. However, although the mainstream CodeLMs have demonstrated preliminary effectiveness in CLB detection, they are still primarily used in research settings and are not yet ready for real-world applications. We then explored factors affecting the performance of CodeLMs in CLB detection. The results show that increasing dataset size improves CodeLM performance, while the impact of token sequence length varies across CodeLMs. Some CodeLMs perform better with a shorter token sequence length, while others perform better with a longer token sequence length. Finally, we investigated the impact of code comments on model performance in CLB detection. Experimental results indicate that code comments affect CodeLMs differently - some CodeLMs perform better with commented code, some CodeLMs show no significant performance change, and others perform worse.

For future work, we will advance research on CLB detection along several key directions.
First, we intend to conduct a more in-depth analysis of the CLB dataset itself. This includes investigating the characteristics of CLBs, such as common interaction patterns across language boundaries, as well as analyzing CLB fix characteristics to better understand how CLBs are resolved in practice. Such analyses will provide deeper insights into the nature of CLBs and help inform the design of more effective detection and repair approaches. 
Second, we plan to expand our dataset beyond Python and Java to include more PLs. This will enable more comprehensive CLB studies and improve the generalizability of detection models. Additionally, since our dataset includes bugs related to high-severity security vulnerabilities, we aim to explore cross-language security vulnerability detection in future research.
Third, while this study focused on relatively small CodeLMs (no larger than 7B), we plan to investigate the performance of large CodeLMs in CLB detection. We also intend to conduct more in-depth analysis to better understand the factors that influence CLB detection performance, such as model architecture and language-specific features. 
Finally, we will examine the effect of longer input sequences and larger datasets on model performance. In this study, input token sequence length was limited to 512, which may have restricted the CodeLM’s ability to capture long-range dependencies. We plan to expand our dataset in order to study the impact of dataset size on CLB detection.

\section*{Data Availability}
The replication package of this work has been made available at \cite{dataset}.

\section*{Acknowledgments}
This work was funded by the National Natural Science Foundation of China under Grant Nos. 62176099 and 92582203, and the Major Science and Technology Project of Hubei Province under Grant No. 2024BAA008. The numerical calculations in this paper have been done on the supercomputing system in the Supercomputing Center of Wuhan University.

\bibliographystyle{ACM-Reference-Format}
\bibliography{references}

\balance

\end{document}